\documentclass[english,twocolumn,prl,amsmath,amssymb,floatfix]{revtex4-1}
\pdfpageattr {/Group << /S /Transparency /I true /CS /DeviceRGB>>}
\usepackage[T1]{fontenc}
\usepackage[latin9]{inputenc}
\usepackage{dcolumn}
\usepackage{bm}
\usepackage{graphicx}
\usepackage{color}
\usepackage{babel}
\usepackage{amsfonts}
\usepackage{mathtools}
\usepackage{slashed}
\usepackage{enumerate}
\usepackage{braket}
\usepackage[multiple]{footmisc}

\usepackage{tikz}
\usetikzlibrary{calc,shapes.arrows,shapes.geometric}
\begin{document}


\title{The Mott transition as a topological phase transition}

\author{Sudeshna Sen$^*$}
\author{Patrick J. Wong$^*$}
\author{Andrew K. Mitchell}
\affiliation{School of Physics, University College Dublin, Belfield, Dublin 4, Ireland}


\begin{abstract}
\noindent We show that the Mott metal-insulator transition in the standard one-band Hubbard model can be understood as a topological phase transition. 
Our approach is inspired by the observation that the mid-gap pole in the self-energy of a Mott insulator resembles the spectral pole of the localized surface state in a topological insulator.  
We use NRG-DMFT to solve the infinite-dimensional Hubbard model, and represent the resulting local self-energy in terms of the boundary Green's function of an auxiliary tight-binding chain without interactions. The auxiliary system is of generalized SSH model type; the Mott transition corresponds to a dissociation of domain walls. 
\end{abstract}
\maketitle


The Mott transition is a classic paradigm in the physics of strongly correlated electron systems, where electronic interactions drive a metal-insulator phase transition~\cite{hubbard1963electron,Imada_RMP1998,mott1974metal}. In a Mott insulator (MI), the strong local Coulomb repulsion localizes electrons, opening a charge gap to single-particle excitations and suppressing transport.

Although most MIs are accompanied by magnetic order at low temperatures, yielding a symmetry-broken superlattice structure~\cite{Imada_RMP1998}, this is not an essential requirement~\cite{mott1974metal,dobrosavljevic2012introduction,Pustogow_2018}. The one-band Hubbard model on the Bethe lattice is the simplest model describing the Mott transition to a paramagnetic MI, and can be solved numerically exactly using dynamical mean-field theory (DMFT)~\cite{DMFT_1996_review,bulla1999}.  
The insulating properties of the MI cannot be understood on the single-particle level; all nontrivial physics is contained in the interaction self-energy~\cite{bulla1999,georges_kotliar_comment,logan2015mott}.
Throughout the insulating phase, the MI self-energy features a mid-gap pole. In the metallic Fermi liquid (FL) phase, Landau damping sets in at low energies. Close to the Mott transition, the FL self-energy develops a double-peak structure responsible for the pre-formed spectral gap, separating the central quasiparticle resonance in the density of states from the high energy Hubbard bands. Importantly, the Mott transition from FL to MI arises without the gap between the Hubbard bands closing. At particle-hole symmetry (half filling), the self-energy peaks sharpen and coalesce to form a single Mott pole pinned at zero energy~\cite{bulla1999,georges_kotliar_comment,logan2015mott}.

MIs contrast to standard band insulators, where the non-interacting band structure is already gapped due to the specific periodic structure of the real-space lattice. Indeed, the topology of the band structure of non-interacting systems plays an important role~\cite{hasankane,RevModPhys.83.1057,moore2010birth}. In particular, topological insulators constitute distinct phases of matter, characterized by robust metallic states localized at boundaries, or at interfaces with trivial insulators~\cite{yang2013topologicalrobustedge,konig2007quantum,xiao2011interface}. Topological phase transitions typically involve bulk gap closing without symmetry breaking, and are characterized by the discrete change in a topological invariant~\cite{moore_balents_topo_invariant}. However, for interacting systems the standard topological classification breaks down~\cite{PhysRevB.81.134509,*PhysRevB.83.075103,PhysRevB.92.125104,pole_expansion_SE}. The effect of including electronic interactions in systems with topologically nontrivial single-particle band structures is the focus of active research~\cite{interactingreview,*irkhin2019modern,amaricci2019,
*amaricci2017edge,*amaricci2015first,*amaricci2016,Hohenadler_2013_review,
tqpt_2014_ed,pizarro2020deconfinement,PhysRevB.85.125113,*kawakami2012,sangiovanni,topologicalmottins,pramod2016,
shenoy,ishida,bijelic2018suppression,*PhysRevLett.112.196404,*PhysRevB.87.085134,magnifico2019symmetry}.

Recently, the violation of Luttinger's theorem in correlated materials has been connected to the emergence of topological order~\cite{wu2018pseudogap,scheurer2018topological,sachdev2018topological,vishwanath,osborne2020topological,Senthil_Sachdev_Vojta,irkhin2019topological,mukherjee2018scaling,oshikawa_LT_topology}. Although Luttinger's theorem is satisfied throughout the FL phase of the Hubbard model due to the vanishing of the Luttinger integral \cite{philip_phillips,dzyalonshkii2003,philip2,oshikawa_LT_topology,logan2015mott}, it is violated in a MI \cite{philip_phillips,rosch2007breakdown,logan2015mott}. Importantly, the Luttinger integral takes a universal finite value throughout the MI phase \cite{logan2015mott}, suggesting that it may play the role of a topological invariant, and that topological information is contained in the interaction self-energy.

\begin{figure}[htp!]
\includegraphics[]{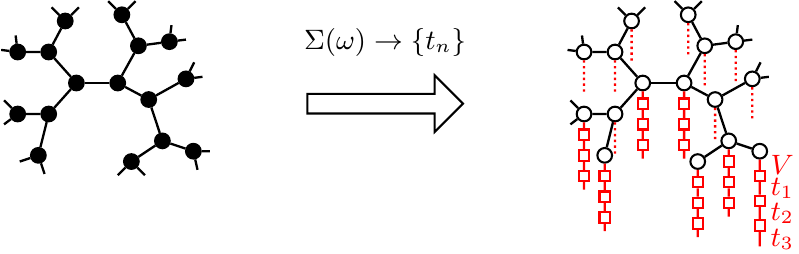}
  \caption{Mapping from the Hubbard model (left) to a~fully non-interacting system (right) in which physical degrees of freedom ($\circ$) couple to auxiliary tight-binding chains ($\square$). }
  \label{fig:schematic}
\end{figure}

In this Rapid Communication, we uncover a hidden topology in the self-energy of the standard one-band paramagnetic Hubbard model in infinite dimensions. Specifically, we show that the rich many-body features of the Mott transition can be interpreted in terms of topological properties of an auxiliary non-interacting system coupled to the physical lattice degrees of freedom.
The original interacting lattice system is mapped onto a completely non-interacting one; the self-energy dynamics are provided by coupling to fictitious degrees of freedom of an auxiliary system, see Fig.~\ref{fig:schematic}. We use NRG-DMFT~\cite{DMFT_1996_review,bulla1999,weichselbaum2007sum} to calculate the zero-temperature local lattice self-energy numerically exactly, perform the exact mapping to an auxiliary tight-binding chain coupled to each physical lattice site, and analyze their topological properties across the Mott transition. The auxiliary chains are found to be of generalized Su-Schreiffer-Heeger \cite{ssh,*shortcourse} (SSH) model type, with the MI being the topologically nontrivial phase. The double peak structure of the self-energy in the topologically trivial FL phase corresponds to an SSH chain with additional domain walls. In each regime, we construct simple effective models to describe the emergent physics.

\emph{Models and mappings.--}
To uncover the topological features of the Mott transition in their simplest form, we focus on the one-band Hubbard model (Fig.~\ref{fig:schematic}, left), 
\begin{equation}
\label{eq:hubb}
H_{\rm{latt}} = H_{\rm{band}}+ H_{\rm{int}}  = \tilde{t} \sum_{\langle i,j \rangle,\sigma} c_{i\sigma}^{\dagger}c_{j\sigma}^{\phantom{\dagger}} + U\sum_{i}c_{i\uparrow}^{\dagger}c_{i\uparrow}^{\phantom{\dagger}}c_{i\downarrow}^{\dagger}c_{i\downarrow}^{\phantom{\dagger}}\;,
\end{equation}
where $\langle i,j \rangle$ denotes nearest neighbours on the Bethe lattice. In the limit of infinite lattice coordination $N\rightarrow \infty$ (considered hereafter), the self-energy $\Sigma(\omega)$ becomes purely local \cite{DMFT_1996_review} such that $G(\omega)=1/ [\omega^{+}-\Sigma(\omega)-t^2 G(\omega)]$, 
where $\omega^{+}=\omega+i0^{+}$, $t=\tilde{t}\sqrt{N}$, and $G(\omega)$ is the retarded lattice Green's function.
We use NRG-DMFT \cite{bulla1999,weichselbaum2007sum} to determine $\Sigma(\omega)$ at $T=0$ across the Mott transition.

Since the self-energy is analytic and causal, it may be replaced by a hybridization $\Sigma(\omega) \equiv \Delta_{0}(\omega)$ to auxiliary (`ghost') degrees of freedom described by some non-interacting $H_{\rm{aux}}$.  
The full single-particle dynamics of Eq.~\ref{eq:hubb} can therefore be reproduced by replacing $H_{\rm{int}}\rightarrow H_{\rm{aux}}+H_{\rm{hyb}}$. 
Specifically, we take $H_{\rm{aux}}$ to be non-interacting semi-infinite tight-binding chains,
\begin{equation}
\label{eq:Haux}
H_{\rm{aux}}=\sum_{i,\sigma} \sum_{n=1}^{\infty} e_n f_{i\sigma,n}^{\dagger}f_{i\sigma,n}^{\phantom{\dagger}} + t_n \left(f_{i\sigma,n}^{\dagger}f_{i\sigma,n+1}^{\phantom{\dagger}}+\rm{H.c.}\right) \;,
\end{equation}
coupled at one end to the physical lattice degrees of freedom, $H_{\rm{hyb}}=V\sum_{i,\sigma} ( c_{i\sigma}^{\dagger}f_{i \sigma,1}^{\phantom{\dagger}}+f_{i\sigma,1}^{\dagger}c_{i\sigma}^{\phantom{\dagger}} )$, Fig.~\ref{fig:schematic} (right).


\emph{Continued fraction expansion.--} 
With $H_{\rm{aux}}$ in the form of a linear chain, $\Delta_0(\omega)$ can be expressed as a continued fraction using $\Delta_{n}(\omega)=t_n^2/[\omega^{+}-e_{n+1}-\Delta_{n+1}(\omega)]$, where $t_0=V$.   
The set of chain parameters $\{t_n\}$ and $\{e_n\}$ in Eq.~\ref{eq:Haux} for a given input self-energy $\Sigma(\omega)$ is uniquely determined using this recursion for $\Delta_n$ (initialized by $\Delta_0=\Sigma$), together with the identities $t_n^2=-\frac{1}{\pi}{\rm{Im}}\int d\omega~\Delta_{n}(\omega)$ and $e_{n+1}=-\frac{1}{\pi t_n^2}{\rm{Im}}\int d\omega~\omega\Delta_{n}(\omega)$. We impose a high-energy cutoff $D$ such that  ${\rm{Im}}\Sigma(\omega)\propto \theta(D-|\omega|)$ \footnote{Results presented are insensitive to the choice of $D$}. 
The mapping is efficient, numerically stable and accurate, although care must be taken with poles in $\Delta_n$ \cite{SM}.

We now focus on the particle-hole symmetric (half-filled) case $\mu=U/2$, where $\text{Im}\Sigma(\omega)=\text{Im}\Sigma(-\omega)$ and so  
$e_n=0$ for all sites of the auxiliary chain.  

\begin{figure}[htp!]
\includegraphics[width=1.0\columnwidth]{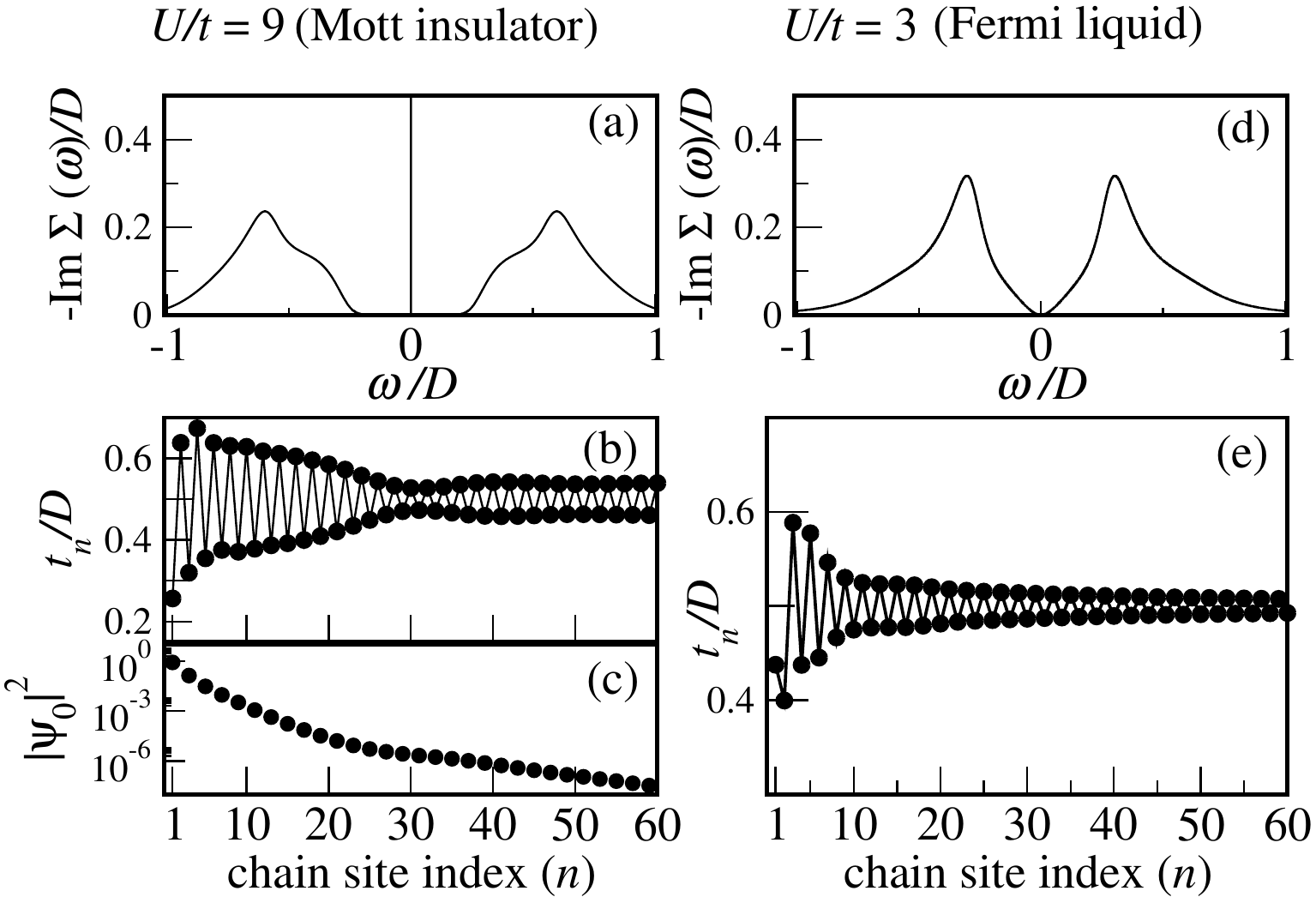}
  \caption{Lattice self-energy at $T=0$ obtained from NRG-DMFT [panels (a,d)] and corresponding $t_n$ of the auxiliary chain [panels (b,e)]. Left panels show results for the MI ($U/t=9$, $D=4$): the hard gap in $\rm{Im}\Sigma(\omega)$ and the Mott pole at $\omega=0$ produce an SSH-type chain in the topological phase, hosting an exponentially-localized boundary zero mode, panel (c). Right panels show the metallic FL ($U/t=3$, $D=3$): the low-energy $\omega^2$ psuedogap in $\rm{Im}\Sigma(\omega)$ produces a generalized SSH chain with $1/n$ decay, in the trivial phase.}
  \label{fig:2}
\end{figure}

\emph{Mott insulator.--}
For interaction strength $U>U_c$, the Hubbard model Eq.~\ref{eq:hubb} describes a MI, with two Hubbard bands separated by a hard spectral gap of width $2\delta$. The corresponding self-energy at zero temperature is shown in Fig.~\ref{fig:2}(a), obtained by NRG-DMFT for $U/t=9$. The imaginary part of the self-energy features a mid-gap `Mott pole' throughout the MI phase, pinned at $\omega=0$ (and with finite weight at the transition).  

Mapping to the auxiliary non-interacting chain, Eq.~\ref{eq:Haux}, leads to a model of modified SSH type -- see Fig.~\ref{fig:2}(b). In particular, the hard gap in $\rm{Im}\Sigma(\omega)$ generates an alternating sequence of $t_n$ in $H_{\rm{aux}}$ at large distances from the physical degrees of freedom,
\begin{equation}
    \label{eq:tn_MI}
    t_n ~~ \overset{n \delta/D \gg 1}{\sim} ~~ \tfrac{1}{2}[D+(-1)^n\delta] \qquad: \rm{MI}
\end{equation}
In the MI phase, the auxiliary chain parameters are alternating for all $n$, \emph{starting from a weak bond} ($t_1<t_2$). It is this feature that produces the Mott mid-gap pole at $\omega=0$. Additional structure in the Hubbard bands merely gives rise to transient structure in the $t_n$ for small $n$, but importantly the parity of the alternation, $t_{2n-1}/t_{2n}<1$, is preserved for all $n$ [see Fig.~\ref{fig:2}(b)].

The SSH model in its topological phase (Eq.~\ref{eq:Haux} with $t_n$ given by Eq.~\ref{eq:tn_MI} for all $n\ge 1$) hosts an exponentially-localized boundary zero-mode that is robust to parity-preserving perturbations~\cite{ssh,*shortcourse}. Similarly, the zero-energy Mott pole corresponds to a robust and exponentially-localized state living at the end of the auxiliary chain (on its boundary with the physical degrees of freedom of the original lattice). This can be readily seen from the transfer matrix method, which gives the wavefunction amplitude of the zero-energy state at odd sites $(2n-1)$ \\of $H_{\rm{aux}}$ as $|\psi_0(2n-1)|^2\sim \prod_{x=1}^{n} t_{2x-1}/t_{2x}$, which at large $n$ decays exponentially as $\exp(-n/\xi)$ with $\xi \approx D/2\delta$ for small $\delta$ 
(while $|\psi_0(2n)|^2=0$ for all $n$)~\cite{ssh,*shortcourse}. The boundary-localized nature of this zero-mode state is confirmed by exact diagonalization of $H_{\rm{aux}}$, see Fig.~\ref{fig:2}(c).


\emph{Metallic FL phase.--}
For $U<U_c$, Eq.~\ref{eq:hubb} describes a correlated metal, with low-energy FL properties characterized by a quadratic dependence of the self-energy, $-t\rm{Im}\Sigma(\omega\rightarrow 0) \sim (\omega/Z)^2$, in terms of the quasiparticle weight $Z$. In Fig.~\ref{fig:2}(d) we plot the $T=0$ self-energy deep in the FL phase, obtained by NRG-DMFT for $U/t=3$.
We obtain a distinctive form for the auxiliary chain hopping parameters from the continued fraction expansion, arising due to the low-energy pseudogap in $\rm{Im}\Sigma(\omega)$, 
\begin{equation}
    \label{eq:tn_FL}
    t_n^2 ~~\overset{nZ\gg 1}{\sim}~~ \frac{D^2}{4} \left [1-\frac{r}{n+d}(-1)^n \right ] \qquad: \rm{FL}
\end{equation}
where $r=2$ is the exponent of the low-energy spectral power-law, and $d\sim 1/Z$. Eq.~\ref{eq:Haux} with hopping parameters $t_n$ given by Eq.~\ref{eq:tn_FL} generalizes the standard hard-gapped SSH model to the pseudogapped case: the alternating sequence of $t_n$ again has a definite parity, but with a decaying $1/n$ envelope. Since $t_{2n-1}/t_{2n}>1$ for all $n$ (the chain starting this time from a \emph{strong} bond), the analogous SSH model would be in its trivial phase; likewise here, the FL phase of the Hubbard model may be regarded as trivial. There is no localized boundary state of the auxiliary chain in the FL phase.


\emph{Vicinity of transition.--}
Deep in either MI or FL phases of the Hubbard model, the auxiliary chains are of generalized SSH model type, with the MI being topologically nontrivial. A robust and exponentially-localized zero-energy state lives on the boundary between the auxiliary and physical systems throughout the MI phase, corresponding to the Mott pole. However, richer physics is observed on approaching the Mott transition from the FL phase. In particular, the Mott transition occurs \emph{without} bulk gap closing of the Hubbard bands (unusual for a topological phase transition). What is the mechanism for the transition between the trivial FL and the topological MI in terms of the auxiliary chains?

In the vicinity of the transition on the FL side, the self-energy develops a preformed gap, inside which are peaks located at $\pm \omega_p$ with $\omega_p\propto t\sqrt{Z}$, while quadratic `pseudogap' behaviour sets in on the lowest energy scales $|\omega|\ll \omega_p$ \cite{bulla1999,georges_kotliar_comment,logan2015mott}. The transition corresponds to $Z\rightarrow 0$. Before performing the exact mapping $\Sigma(\omega)\rightarrow \{t_n\}$ numerically, we consider the evolution of chain parameters for a simpler toy system mimicking the Mott transition: two mid-gap spectral poles merging to one.

To do this, we consider the general problem of determining the chain parameters $t_n$ for a composite spectrum $A(\omega)=\tfrac{1}{\mathcal{N}}\sum_i w_i A_i(\omega)$, with $\mathcal{N}=\sum_i w_i$. Although spectral elements are simply additive, the composition rule for the $t_n$ is highly non-linear. To make progress we note that spectral moments are additive, $\mu_{k}=\tfrac{1}{\mathcal{N}}\sum_i w_i \mu_{i,k}$ with $\mu_{i,k} = \int d\omega~ \omega^{k} A_i(\omega)$, and use the moment expansion~\cite{vishwanath1994g} of the chain parameters $t_n^2=X_{n}(n)$, where 
\begin{eqnarray}
\label{eq:ME}
X_{k}(n) = \frac{X_{k}(n-1)}{t_{n-1}^2}-\frac{X_{k-1}(n-2)}{t_{n-2}^2} \;, 
\end{eqnarray}
with $X_{k}(0) = \mu_{2k}$, $X_{k}(-1)=0$ and $t_{-1}^2=t_0^2=1$.

\begin{figure}[t]
  \centering
\includegraphics[width=1.0\columnwidth]{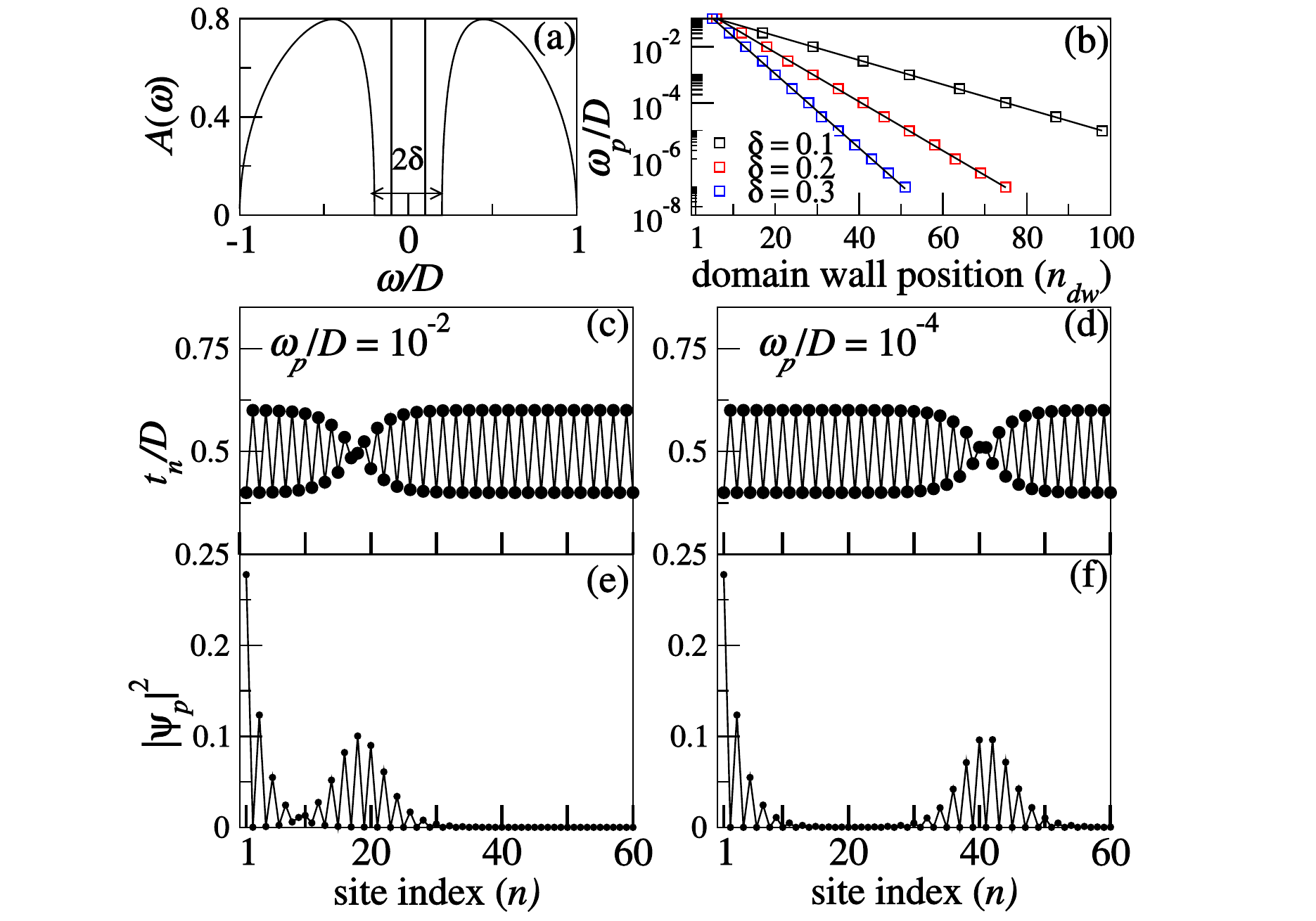}
  \caption{Modified SSH model with two poles at $\pm\omega_p$ inside a gap of width $2\delta$ [spectral function illustrated in panel (a)]. Chain parameters $t_n$ presented in panels (c) and (d) for $\omega_p/D=10^{-2}$ and $10^{-4}$ with common $\delta/D=0.2$, showing a domain wall at $n_{\rm{dw}}$. States localized at the boundary and the domain wall hybridize and gap out to give exact eigenstates with energies $\pm\omega_p$, panels (e) and (f). The domain wall position [panel (b), points] follows $\omega_p\sim D \exp(-n_{\rm{dw}} \delta/D)$ (lines).  
  }
   \label{fig:ssh}
\end{figure}

Analysis of the equations shows that adding a zero-energy pole to the boundary spectral function of the SSH model in the trivial phase flips the parity of the corresponding $t_n$ (the first coupling of the chain swaps from a strong to a weak bond), yielding the topological SSH model, Eq.~\ref{eq:tn_MI}, as expected. What change in $t_n$ results from adding \emph{two} poles at $\pm\omega_p$ to the trivial SSH spectrum, as depicted in Fig.~\ref{fig:ssh}(a)?

Figs.~\ref{fig:ssh} (c,d) show the chain parameters $t_n$ for $\omega_p/D=10^{-2}$ and $10^{-4}$. At large $n$, the chain remains in the trivial SSH phase. However, a domain wall appears at $n_{\rm{dw}}$ where the parity of the alternation flips; the chain for $1<n<n_{\rm{dw}}$ is therefore in the \emph{topological} phase of the SSH model (starting at $n=1$ from a \emph{weak} bond). This produces \emph{two} localized states -- one at the boundary ($n=1$), and the other pinned at the domain wall ($n=n_{\rm{dw}}$), which hybridize and gap out to produce two states at energies $\pm\omega_p$. Since these are topological states and exponentially localized, the hybridization is exponentially small in the real-space separation between them along the chain, and we find $\omega_p\sim D \exp(-n_{\rm{dw}} \delta/D)$, see panel (b). This physical picture is confirmed by 
examining the exact eigenstates $\psi_p$ with energy $\omega_p$ satisfying $H_{\rm{aux}}\psi_p=\omega_p\psi_p$, plotted in panels (e,f).

The Mott transition as $U\rightarrow U_c^{-}$ is characterized by $\omega_p\rightarrow 0$. In terms of the auxiliary chain, a \emph{pair} of topological defects forms at the boundary when deep in the FL phase. One of these separates and moves down the chain as the transition is approached. As $U\rightarrow U_c^{-}$ then $\omega_p\rightarrow 0$, and $n_{\rm{dw}}\rightarrow \infty$. At the transition itself, the two poles coalesce into the single Mott pole, and the chain is left with a single topological defect state at the boundary. This mechanism is reminiscent of the vortex-pair dissociation in the Kosterlitz-Thouless transition~\cite{kosterlitz1972long,*kosterlitz1973ordering}. The topological transition occurs without bulk gap closing. 

\begin{figure}[t]
  \centering
\includegraphics[width=1.0\columnwidth]{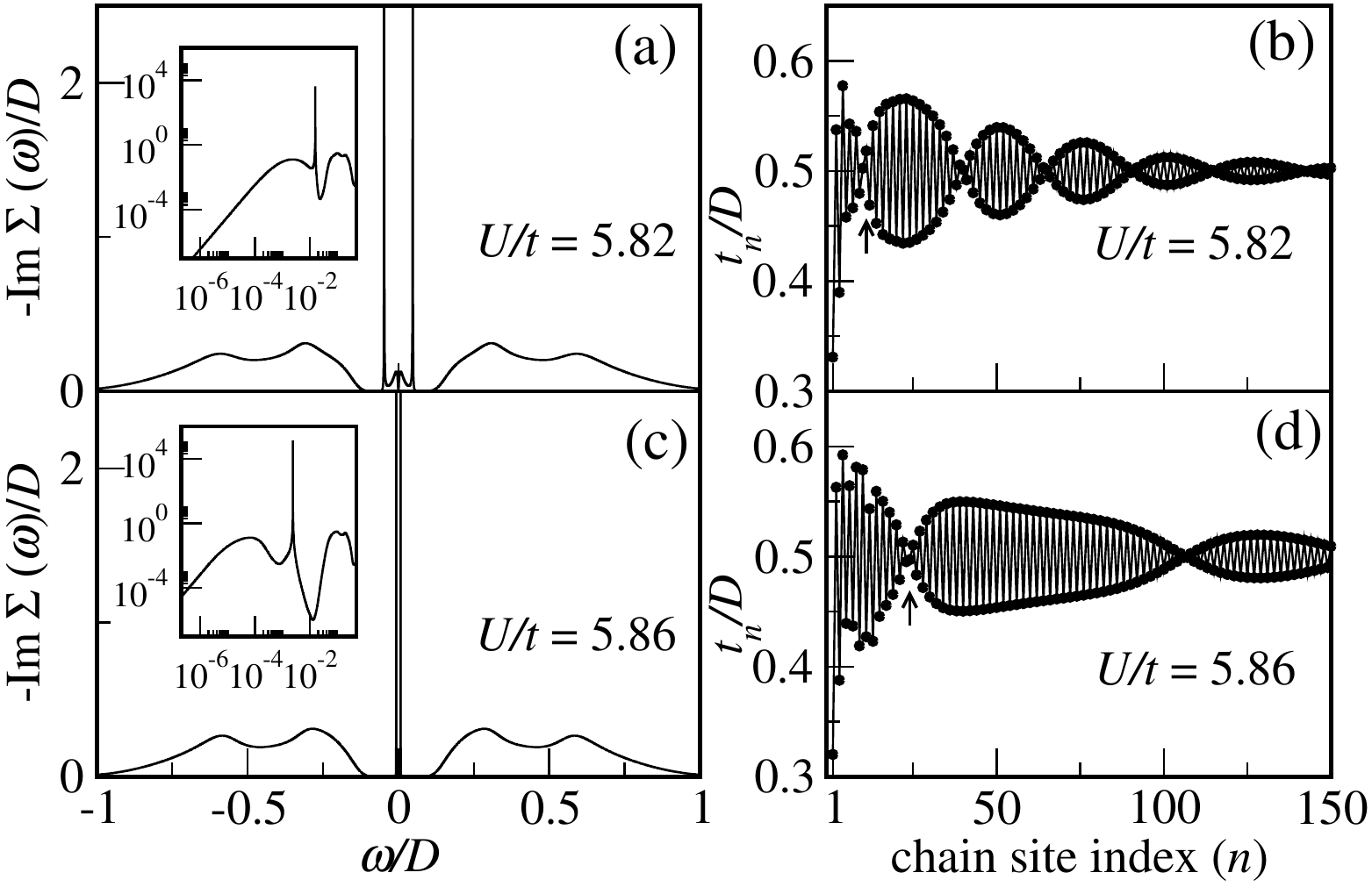}
  \caption{Self-energy $-{\rm{Im}}\Sigma(\omega)$ from NRG-DMFT (a,c) and corresponding auxiliary chain parameters $t_n$ (b,d) close to the Mott transition in the FL phase at $T=0$. Top panels for $U/t=5.82$; lower panels for $U/t=5.86$ (both with $D=3$). Self-energy peaks of finite width centred on $\pm \omega_p$ produce a generalized SSH chain with periodic domain wall structure. Low-energy $\omega^2$ behaviour of the self-energy manifests as long-distance $(-1)^n/n$ behaviour in the chains. As the transition is approached, the self-energy peaks sharpen into poles and $\omega_p\rightarrow 0$; correspondingly, the location of the first domain wall moves out, and the beating period increases, leaving a single boundary-localized topological state in the MI.}
  \label{fig:exactFL}
\end{figure}

The behaviour of the auxiliary chains for the actual Hubbard model is of course more complex than that of the above toy model. In particular, the true self-energy $\Sigma(\omega)$ is not completely hard-gapped in the FL phase, but features a low-energy quadratic pseudogap. Including this leads to alternating $t_n$ with a $1/n$ envelope as per Eq.~\ref{eq:tn_FL}. Another key difference is that the peaks in the self-energy close to the transition are not delta-functions but have finite width. For the auxiliary chains, these peaks can be viewed as narrow \emph{bands} of hybridizing topological states produced by a periodic structure of domain walls, as shown in the Supplemental Material \cite{SM}. One therefore expects a beating pattern in the chain parameters.

All these expected features are seen in the exact results for the self-energy and corresponding chain parameters close to the transition, shown in Fig.~\ref{fig:exactFL}. In particular, the chains start from a weak bond (giving a localized boundary state); the position of the first domain wall moves to larger distances as $\omega_p$ becomes smaller nearer the transition; the period of the beating becomes longer as the self-energy peaks become sharper; and the alternation in $t_n$ attenuates as $1/n$ at long distances.

\begin{figure}[t]
  \centering
\includegraphics[width=1.0\columnwidth]{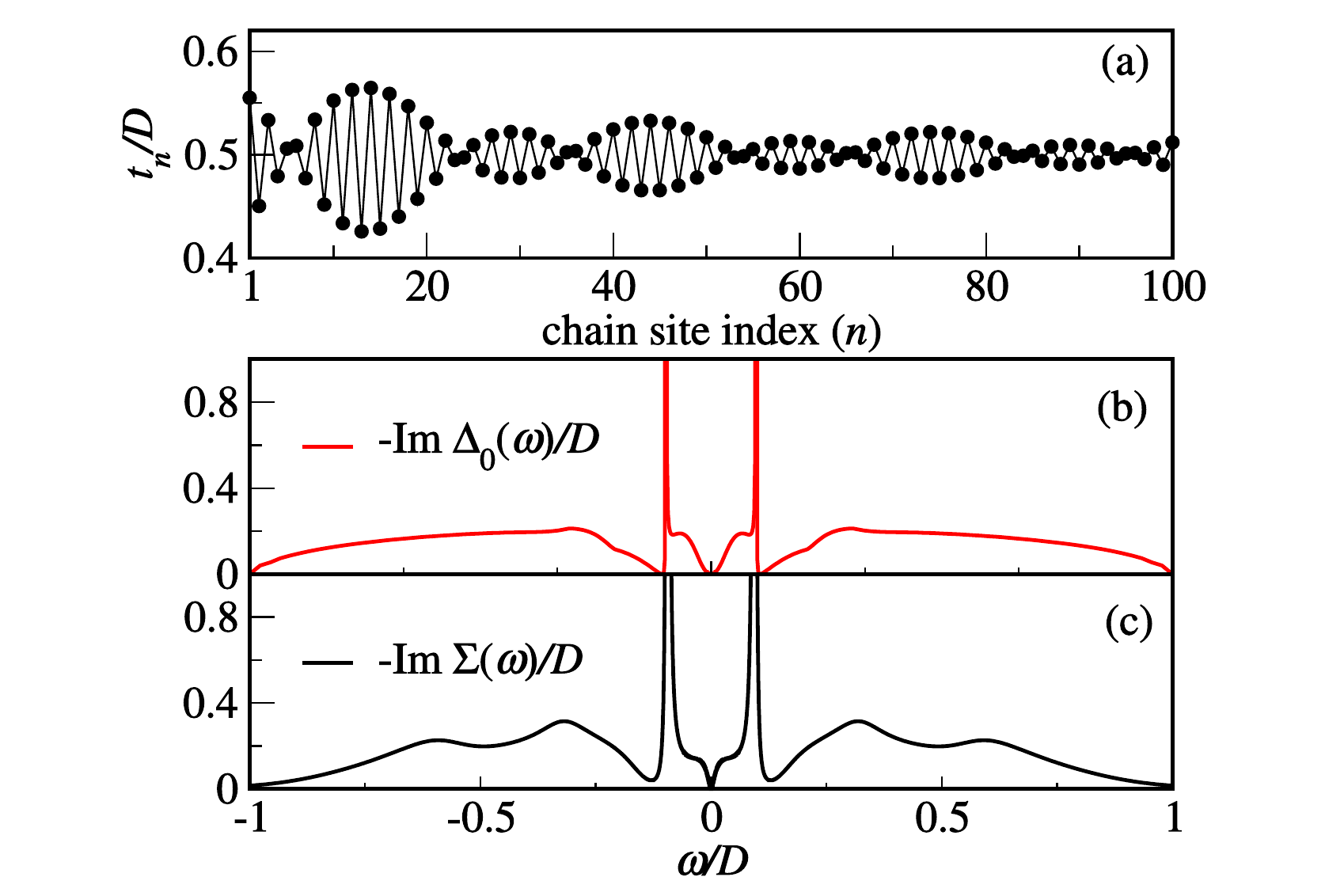}
  \caption{Auxiliary chain parameters $t_n$ of the toy model Eq.~\ref{eq:toy}, with parameters $\beta=3$, $d=15$, $\phi=0.1$ and $\lambda=30$ [panel (a)]. The resulting $\Delta_0(\omega)$ [panel (b)] is in good agreement with the true lattice self-energy of the Hubbard model for $U/t=5.6$, $D=3$ [panel (c)].
  }
  \label{fig:toy}
\end{figure}

Combining these insights, we propose a simple toy model that approximates all of the qualitative features of the true lattice self-energy throughout the FL phase:
\begin{equation}
    \label{eq:toy}
    t_n^2 = \frac{D^2}{4}\left [1-\frac{2}{n+d}(-1)^n \right ]\times [1-\beta\cos(2\pi n/\lambda +\phi) ]\;.
\end{equation}
A representative example is shown in Fig.~\ref{fig:toy}(a), where we have fit the parameters of Eq.~\ref{eq:toy} to best match $\Sigma(\omega)$ from NRG-DMFT [panel (c)] with $\Delta_{0}(\omega)$ of the toy model [panel (b)]. The transition is approached as $\lambda, d\rightarrow \infty$.


\emph{Particle-hole asymmetry.--}
We briefly comment on the physics away from particle-hole (\textit{ph}) symmetry, $\eta=1-2\mu/U\ne 0$. 
Throughout the MI phase, the Mott pole resides inside the hard gap between Hubbard bands, but is no longer at zero energy. The resulting auxiliary chain potentials $e_n$ then become finite. We have confirmed in this case that the auxiliary chain state to which the Mott pole corresponds is still exponentially localized on the boundary, and is robust to physical perturbations (provided one remains in the MI phase). Furthermore, the analysis of Ref.~\onlinecite{PhysRevB.89.085111} can be applied to the auxiliary chain. We again find that the MI is topologically non-trivial for $\eta \ne 0$ (while FL is trivial). Everything is continuously connected to the \textit{ph}-symmetric limit $\eta\rightarrow 0$. Further details and explicit calculations for $\eta=1/4$ are presented in the Supplemental Material \cite{SM}. A full discussion will appear elsewhere.


\emph{Topological invariant.--}
A recent paper by Logan and Galpin \cite{logan2015mott} shows for the Hubbard model Eq.~\ref{eq:hubb} at $T=0$ that the Luttinger integral takes distinct constant values in the FL and MI phases for any $\eta\ne 0$ 
\footnote{$\eta=0$ is a special point at $T=0$ where the authors of Ref.~\cite{logan2015mott} find $I_{\text{L}}=0$. This appears to be an order-of-limits issue and $I_{\text{L}}=1$ in the MI is expected if $\eta\rightarrow 0$ is taken before $T\rightarrow 0$ (M.~Galpin, Private Communication).}\footnote{For a detailed discussion of the Luttinger integral in non-Fermi liquid phases, see: D.~E.~Logan, A.~P.~Tucker, and M.~R.~Galpin, Phys.~Rev.~B \textbf{90}, 075150 (2014).},
\begin{eqnarray}
\label{eq:lutt}
I_{\rm{L}}=\frac{2}{\pi}{\rm{Im}}\int_{-\infty}^{0} d\omega ~G(\omega) \frac{d \Sigma(\omega)}{d\omega} ~ = ~ \begin{cases}
0& {\rm{:~FL}} \;,\\
1& {\rm{:~MI}} \;.
\end{cases}
\end{eqnarray}
The finite value of $I_{\rm{L}}$ for the generic MI can be traced to the Mott pole, which we identified in this work as the topological feature of the MI. 
Since the evolution of the self-energy with interaction strength drives the Mott transition, the Luttinger integral is a natural quantity to characterize the distinct topology of the FL and MI phases, and may be regarded as a topological invariant.


\emph{Conclusions.--}
We present an interpretation of the classic Mott transition in the infinite-dimensional one-band Hubbard model as a topological phase transition. The lattice self-energy, determined here by NRG-DMFT, is mapped to an auxiliary tight-binding chain, which is found to be of generalized SSH model type. The MI is the topological phase, with a boundary-localized state corresponding to the Mott pole. The transition from FL to MI involves domain wall dissociation. 

We argue that any system with such a pole in its local self-energy may be regarded as topological.
The analysis could also be extended to multi-band models, where the auxiliary chains become multi-legged ladders. We speculate that a superconducting Hubbard model may map to auxiliary Kitaev chains involving Majoranas. 
For a fully momentum-dependent self-energy of a $D$-dimensional lattice, the mapping generalizes to an auxiliary \emph{lattice} in $D+1$ dimensions; for a MI, the auxiliary lattice may be a topological insulator with a localized boundary state.


\begin{acknowledgments}
\emph{Acknowledgments.--} This paper is dedicated to the memory of Mark Jarrell. We acknowledge discussions with Siddhartha Lal and N. S. Vidhyadhiraja.
We acknowledge funding from the Irish Research Council Laureate Awards 2017/2018 through grant IRCLA/2017/169.
\end{acknowledgments}


%


\end{document}


\title{The Mott transition as a topological phase transition:\\
Supplemental Material}
\author{Sudeshna Sen$^*$}
\author{Patrick J. Wong$^*$}
\author{Andrew K. Mitchell}
\affiliation{School of Physics, University College Dublin, Belfield, Dublin 4, Ireland}
\maketitle

This Supplemental Material contains further details on the following topics. Note that sections \ref{s1}--\ref{s5} focus on the particle-hole symmetric case, $\mu=U/2$. In section~\ref{s1} we provide technical details of the continued fraction expansion (CFE) of the Hubbard model self-energy and discuss the analytic structure of auxiliary chain correlation functions. In section~\ref{s2} we demonstrate the veracity of the obtained auxiliary chain hopping parameters $\{t_n\}$, by reproducing the input. In section~\ref{s3} we discuss the `moment expansion' technique, and its advantages and drawbacks in the context of the current work. In section~\ref{s4} we provide additional analysis of the role of the mid-gap peaks on the chain structure. In section~\ref{s5} we discuss an alternative topological signature of the auxiliary chains based on boundary conductance arguments. In section~\ref{sec:ph-asymm} we discuss the auxiliary chain mapping and topological characterization for the Hubbard model in the generic case away from particle-hole symmetry, $\mu\ne U/2$. Finally, the technical details of the NRG and DMFT calculations are presented in Section~\ref{s6}.


\section{Auxiliary chain correlation functions}
\label{s1}
In this section we discuss technical details of the CFE for the mapping of the self-energy $\Sigma(\omega)$ to the auxiliary chain parameters $\{t_n\}$ in the particle-hole symmetric case (half-filling). The $\{t_n\}$ constitute a kind of `genetic code' describing the role of electronic correlations in these systems. We also examine the mathematical structure of the auxiliary hybridization functions $\Delta_n(\omega)$.

The local lattice self-energy $\Sigma(\omega)$ as a function of real frequency $\omega$, is the input for the calculations, and is assumed to be known. In the present work we use NRG-DMFT to obtain it at $T=0$ (see Sec.~\ref{s6}), but in principle any suitable method can be used. The self-energy is identified with a hybridization $\Delta_0(\omega)$ between the physical lattice degrees of freedom and fictitious auxiliary degrees of freedom which provide the same (single-particle) dynamics, $\Delta_0(\omega)=\Sigma(\omega)$. Furthermore, this hybridization can be written in terms of the boundary Green's function of the free auxiliary system -- here a semi-infinite tight-binding chain given by Eq.~2 of the main paper. We write $\Delta_0(\omega)=V^2 G_1(\omega)$, with $V$ the tunneling matrix element between the physical lattice sites and the auxiliary chains appearing in $H_{\rm{hyb}}$, and $G_1(\omega)=\langle\langle f_{i\sigma,1}^{\phantom{\dagger}} ; f_{i\sigma,1}^{\dagger}\rangle\rangle'$ is the retarded Green's function (the prime denotes the isolated $H_{\rm{aux}}$). We define  $\mathcal{A}_n(\omega)=-\tfrac{1}{\pi}{\rm{Im}}~\Delta_n(\omega)$, such that $\int d\omega \mathcal{A}_0(\omega) = V^2$.


\subsection{Continued fraction expansion of self-energy}
The boundary Green's function of the auxiliary chain can be expressed as $G_1(\omega)=1/[\omega^{+}-\Delta_1(\omega)]$ where $\omega^{+}=\omega+i0^{+}$ and $\Delta_1(\omega)$ is the hybridization to the rest of the chain. $\Delta_1(\omega)=t_1^2\tilde{G}_2(\omega)$ can also be expressed in terms of a Green's function. In this case, it is the boundary Green's function of a chain starting at site $n=2$ (i.e., the Green's function at the end of the auxiliary chain with site 1 removed). Following the same logic recursively, we obtain $\tilde{G}_n(\omega)=1/[\omega^{+}-\Delta_{n}(\omega)]$ for $n\ge 1$, where $\Delta_n(\omega)=t_n^2\tilde{G}_{n+1}(\omega)$. Here, $\tilde{G}_{m}(\omega)=\langle\langle f_{i\sigma,m}^{\phantom{\dagger}} ; f_{i\sigma,m}^{\dagger}\rangle\rangle'$ is the Green's function of the auxiliary chain at site $m$ with all sites $<m$ removed. We therefore obtain the continued fraction expansion,
\begin{align}
    \Delta_0(\omega)=\cfrac{V^2}{\omega^+-\cfrac{t_1^2}{\omega^+-\cfrac{t_2^2}{\phantom{a}\ddots}}} \;.
    \label{eq:G_CFE}
\end{align}
Starting from $\Delta_0(\omega)=\Sigma(\omega)$, and using $\int d\omega \mathcal{A}_n(\omega) \equiv -\tfrac{1}{\pi}\int d\omega~{\rm{Im}}\Delta_n(\omega) = t_n^2$, one can recursively determine all $t_n$. Although in practice the recursive procedure is stopped after a finite number of iterations, $N$, note that the $t_n$ determined \emph{en route} are (numerically) exact and are not affected by truncating the sequence at finite $N$ (this is in contrast to methods involving discretization of the spectrum, or the moment expansion discussed below which is strongly nonlinear). For large enough $N$, the asymptotic properties of the chain parameters can be identified, and can be analytically continued, using a chain `terminator'. We observe that thousands of sites of the auxiliary chain must typically be determined to capture low-energy features in the FL phase. In the MI phase, shorter chains of a few hundred sites are usually sufficient to identify the asymptotic chain properties. We emphasize that in all cases an accurate representation of the self-energy requires long (preferably analytically continued) auxiliary chains. Representations involving only a few `ghost' sites are not adequate. 

Finally, we point out that accurate self-energies with high resolution in frequency must be used to recover the auxiliary chain properties discussed in this work. 
\clearpage


\subsection{Auxiliary chain representation of a \\Fermi liquid (FL)}
\label{s1_FL}
In the FL phase, the technical complexity in determining $\{ t_n\}$ numerically is due to the low energy form of the (input) self-energy. In this case, 
\begin{align*}
    \Delta_0(\omega)\overset{\omega\to 0}{\sim} a_0\omega+ib_0\omega^2 \qquad (a_0, b_0< 0) \;.
\end{align*}
\textbf{Step 1:} Calculation of $\Delta_1(\omega):$
\begin{align*}
    \Delta_1(\omega)=t_1^2 \tilde{G}_2(\omega)=\omega^+-1/G_1(\omega).
\end{align*}
Since both the real and imaginary parts of $G_1(\omega)$ is vanishingly small as $\omega\to 0$ and is equal to zero at $\omega=0$, this leads to a non-analyticity in $\Delta_1(\omega=0)$ and hence a singular part in the corresponding Green's function $\tilde{G}_2(\omega)$. We write $\tilde{G}_2(\omega)=\tilde{G}_2^{reg}(\omega)+\tilde{G}_2^{sing}(\omega)$, where $\tilde{G}_2^{reg}(\omega)$ represents the regular (continuum) part, and $\tilde{G}_2^{sing}(\omega)$ represents the singular part. More precisely,
\begin{align*}
    \Delta_1(\omega\to0)=&\omega^+-\frac{t_0^2}{a_0\omega+ib_0\omega^2}\\
    &=\omega^+-\frac{t_0^2}{a_0}\left[\frac{1}{\omega^+}-\frac{ib_0}{a_0+ib_0\omega}\right] \;.
\end{align*}
Therefore we identify 
\begin{align}
\Delta_1^{reg}(\omega\to 0)=\omega+\frac{t_0^2b_0^2\omega}{a_0(a_0^2+b_0^2\omega^2)}+i\frac{t_0^2b_0}{a_0^2+b_0^2\omega^2} \;, \nonumber
\end{align}
such that,
\begin{align}
\Delta_1(\omega\to 0)=\Delta_1^{reg}(0)-\frac{t_0^2}{a_0\omega^+} \;.
\label{eq:sigma1}
\end{align}
The second term in Eq.~\eqref{eq:sigma1} corresponds to a pole in the imaginary part concomitant with a diverging real part of $\Delta_1(\omega)$ at $\omega=0$. Furthermore, this pole resides on top of a background function, $\Delta_1^{reg}(\omega)$, such that $\Delta_1^{reg}(0)=\beta_1$. The residue of this pole is   $\alpha_1=\frac{t_0^2}{|a_0|}$. To fix $t_1^2$ we use the spectral normalization,
\begin{align}
   t_1^2 = \int \mathcal{A}_{1}^{reg}(\omega) d\omega +\alpha_1 \;.
    \label{eq:t1sq}
\end{align}

\noindent\textbf{Step 2:} The low energy spectral behavior of $\Delta_2(\omega)$ is,
\begin{align}
    \Delta_2(\omega)=\omega-\frac{\omega t_1^2}{\omega\Delta_1^{reg}(\omega)+\alpha_1}.
\end{align}
Evaluating the imaginary part,
\begin{align}
{\rm{Im}}\,\Delta_2(\omega)&=-\omega t_1^2 {\rm{Im}}\, \frac{1}{\omega\Delta_1^{reg}(\omega)+\alpha_1}\nonumber \\
    &=\frac{\omega^2 t_1^2 {\rm{Im}}\,\Delta_1^{reg}}{(\omega{\rm{Re}}\,(\Delta_1^{reg})+\alpha_1)^2+(\omega{\rm{Im}}\,(\Delta_1^{reg}))^2}.
    \label{eq:sigma_even_site}
\end{align}
Substituting the respective low energy dependencies of $\Delta_1^{reg}(\omega)$ we find that  ${\rm{Im}}\Delta_2(\omega)\overset{\omega\to0}{\sim} b_2\omega^2$. Similarly, if we evaluate $\rm{Re}\, \Delta_2(\omega\to 0)$ it follows from the presence of a non-zero $\alpha_1$ that $\rm{Re}\, \Delta_2(\omega)\overset{\omega\to0}{\sim} a_2\omega$, just like a Fermi liquid. The advantage of separating the regular and singular part of the odd-site chain hybridizations is thus clear from the structure of Eq.~\eqref{eq:sigma_even_site}, where the information about the underlying pole from the previous iteration is embedded via its weight, and allows us to deal with a regular function numerically.

Based on the above, it is clear that at every odd recursion step, $\Delta_{2n+1}(\omega)$ will have a pole structure similar to $\Delta_1(\omega)$ with a pole weight $\alpha_{2n+1}=\frac{t_{2n}^2}{|a_{2n}|}$ and subsequently the FL character will follow for every even numbered site in the chain, namely, $\Delta_{2n}(\omega)$. In summary, the following recursion relations describe the flow of the low-energy expansion coefficients, 
\begin{align}
&\Delta_{2n}(\omega\to 0)=a_{2n}\omega+ib_{2n}\omega^2 \qquad (a_{2n},b_{2n}<0) \;,\\
&\Delta_{2n+1}(\omega\to 0)=\frac{\alpha_{2n+1}}{\omega^+} + i\beta_{2n+1} \;,\\
& a_{2n}=1-\frac{t_{2n-1}^2}{\alpha_{2n-1}} \qquad;\qquad b_{2n}=\frac{t_{2n-1}^2 \beta_{2n-1}}{\alpha_{2n-1}^2}\;,\\
& \alpha_{2n+1}=\frac{t_{2n}^2}{|a_{2n}|} \qquad\quad~;\qquad \beta_{2n+1}=\frac{t_{2n}^2 b_{2n}}{a_{2n}^2}\,.  
\end{align}
The asymptotic properties of the $\Delta_n(\omega)$ are therefore completely determined by the initialized values $a_0, b_0$ and the $\{t_n\}$ determined by the above CFE.

Note that the above analytic structure is a fingerprint of the FL phase: all ${\rm{Im}}\Delta_n(\omega)$ for \emph{even} $n$ have low-energy quadratic behaviour, while all \emph{odd} $n$ functions have zero-energy poles.

On the practical level of the numerical calculations, for every odd iteration we cut the singular pole feature from $\mathcal{A}_{2n+1}(\omega)$, perform a Kramers-Kr\"onig transformation to obtain the regular real part, and hence $\Delta_{2n+1}^{reg}(\omega)$. The even or odd $t_{n}$ are subsequently evaluated at each iteration from the normalisation of $\mathcal{A}_{n}(\omega)$ as above.


\subsection{Auxiliary chain representation of the \\Mott insulator (MI)}
For a MI, $\Delta_0(\omega)$ is hard-gapped for $\omega\in[-\delta:\delta]$, where $2\delta$ is the size of the Mott gap. Inside the gap resides a zero-energy pole (the `Mott pole'), such that $\Delta_0(\omega\to 0)=\frac{\alpha_0}{\omega^+}$. Based on the analysis for the FL self-energy in the previous subsection, we readily observe that in the MI case, the role of odd and even chain sites is now interchanged. In addition, we note that $\beta=0$ for all (even) $n$ for the MI because the pole is sitting inside the Mott gap and the $b$  coefficients are also zero for all odd $n$. The low-energy asymptotic behaviour and coefficient recursion (for $n\ge1$) follows,
\begin{align}
&\Delta_{2n-1}(\omega\to 0)=a_{2n-1}\omega \;,\\
&\Delta_{2n}(\omega\to 0)=\frac{\alpha_{2n}}{\omega^+}\;,\\
& a_{2n-1}=1-\frac{t_{2n-2}^2}{|\alpha_{2n-2}|} \qquad;\qquad b_{2n-1}=0\;, \\
& \alpha_{2n}=\frac{t_{2n-1}^2}{|a_{2n-1}|} \qquad\qquad~~;\qquad \beta_{2n}=0\,.
\end{align}
Thus the pole structure of the MI $\Delta_n(\omega)$ is reversed with respect to the FL. 
Importantly, the imaginary part of the hybridizations $\Delta_n(\omega)$ is hard gapped for all $n$, but contains a mid-gap zero energy pole for all \emph{even} $n$ (only).


\section{Self-energies reproduced from numerically evaluated $t_n$ sequences}
\label{s2}
Using the the semi-analytic procedure described in section~\ref{s1} we obtain a sequence of the auxiliary chain hopping parameters, $\{t_n\}$. We terminate the recursion after a finite number of steps, $N$ (typically a few thousand). As a check that the derived auxiliary chain does faithfully describe the input self-energy, we numerically evaluate the continued fraction Eq.~\ref{eq:G_CFE}. In practice, we use $\omega^{+}=\omega+i\delta$ with some small but finite $\delta>0$. We also analytically continue the chain beyond step $N$ for a further $~10^6$ sites using the identified asymptotic behaviour. The form of this `terminator' is given by Eq.~4 or Eq.~5 of the main text.

\begin{figure}[t]
  \includegraphics[clip=,scale=0.55]{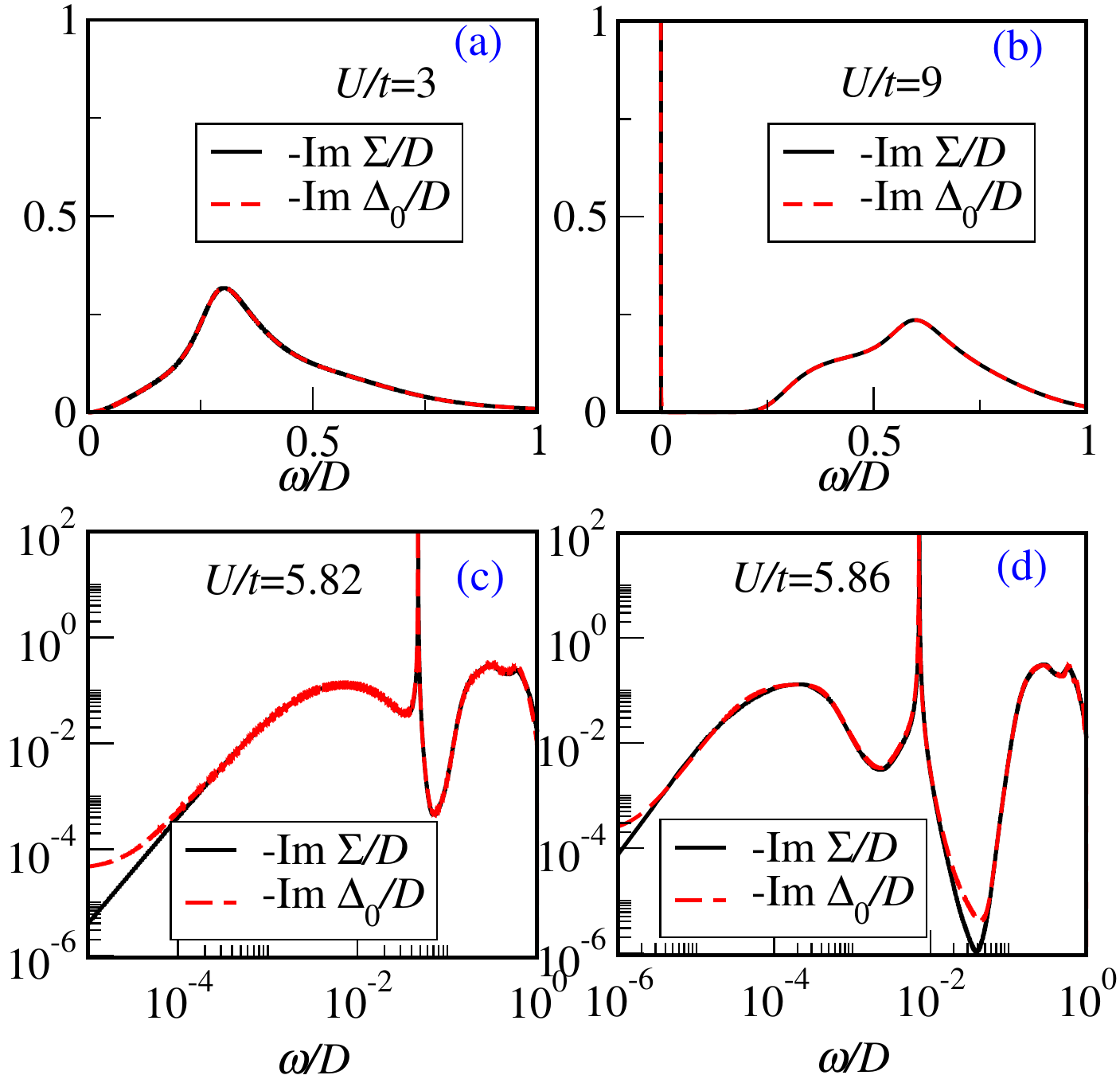}
  \caption{The imaginary part of the $T=0$ Hubbard model self-energy, $-\rm{Im}\, \Sigma(\omega)$ is plotted (black solid lines), calculated using NRG-DMFT on the Bethe lattice with free half-bandwidth $2t$. We compare this with the boundary hybridization function, $\Delta_0(\omega)$ (red dashed line) of the auxiliary chain determined by CFE. The interaction parameters used to obtain $-\rm{Im}\,\Sigma(\omega)$ are: (a) $U/t=3$, (b) $U/t=9$, (c) $U/t=5.82$, (d) $U/t=5.86$. Panel (a) represents a Fermi liquid phase and panel (b) depicts a Mott insulating phase, both being in parameter regimes that are far from the Mott metal-insulator transition. Panels (c, d) represent parameter regimes where the system is close to the Mott transition (approaching from the Fermi liquid side) with sharp pole like features apparent in the self-energy function. $\Delta_0(\omega)$ is calculated using the respective auxiliary chain hopping parameters, $t_n$ (shown in the main text in Figures 2(b,e) and 4(b, d)). For practical purposes we also extrapolate the finite $t_n$ sequence asymptotically using a finite chain terminator to simulate an effectively semi-infinite chain. The excellent agreement between $\Delta_0(\omega)$ and $\Sigma(\omega)$ guarantees a faithful representation of the original self-energy function in terms of the semi-infinite one dimensional auxiliary chain represented by Eq.~\eqref{eq:G_CFE}. The slight discrepancy at extremely low energies is due to the finite broadening factor ($\delta=10^{-6}$ for (a,b,c) and $\delta=10^{-7}$ (d)).}
  \label{fig:S1}
\end{figure}

In Figure~\ref{fig:S1} we compare $-\rm{Im}~\Delta_0(\omega)$ from the CFE with the input $-\rm{Im}\,\Sigma(\omega)$ for different interaction strengths $U$. As can be seen the $t_n$ sequence obtained, excellently reproduces the original Hubbard model self energy, $\Sigma(\omega)$. Any deviations at very small $\omega$ are due to the finite $\delta$ used, and can be systematically pushed to lower $\omega$ by using a smaller $\delta$ together with a longer chain.

\section{Moment expansion technique}
\label{s3}
Given a spectral function, the spectral moments are determined from $\mu_n=\int_{-D}^D \omega^n A(\omega)d\omega$ where $D$ is the half bandwidth. $\mu_0=1$ for a normalised spectral function. Furthermore, for a particle-hole symmetric system, the spectrum follows  $A(\omega)=A(-\omega)$, and only the even moments survive, $\mu_{2n}=0$. We consider this case explicitly in this section (although the method can be generalized). 

For more details the reader is referred to Ref.~\onlinecite{vishwanath1994g} and references therein, where this scheme is discussed on more general grounds. Here we outline only the main equations that were utilised to generate the hopping parameters evaluated in Figure~3 of the main text. Also, here we discuss only the case of particle-hole symmetric spectral function.

As described in Eq.~5 of the main text, given a set of spectral moments, $\lbrace\mu_0,\mu_2,\mu_4,\hdots, \mu_{2N}\rbrace$ it is possible to generate the first $N$ coefficients in the set $\{t_n\}$.

However, it should be noted that while this recursive scheme appears straightforward, it is well known in the literature~\cite{Gaspard_1973,mom_unstable1} that determination of the CFE coefficients via moment expansion is numerically very unstable. 
The spectral moments essentially must be known exactly (obtaining them from numerical integration of the raw spectrum is not sufficient). Furthermore, the recursive calculation must be performed on the computer with arbitrary precision numerics (the scheme breaks down after about 15 steps even with analytically known moments when using standard double-precision numerics).\cite{Gaspard_1973} For our toy models discussed in the main text, where the spectral moments are known exactly, and using arbitrary precision numerics, we could extract reliably the first ~200 chain coefficients before the calculation breaks down and become non-physical.

Therefore, while seemingly appealing, the moment expansion method is not in general of use for determining the auxiliary chain coefficients for the self-energy. We use it for Fig.~3 because we can treat poles and hard gaps cleanly.


\section{Mid-gap peaks as bands of hybridizing topological states}
\label{s4}
In this section we elaborate on the role of the mid-gap peaks in a gapped spectrum using the example of a model system. We extend our analysis represented by Fig~3 of the main text by asking a further question: What change in $\{t_n\}$ results from the consideration of sharp mid-gap peaks centred around $\pm\omega_p$ instead of two poles? Before making connections to the Hubbard model self-energy, we understand this from the perspective of a toy model as discussed in the following.

\begin{figure}[t]
  \includegraphics[width=1.0\columnwidth]{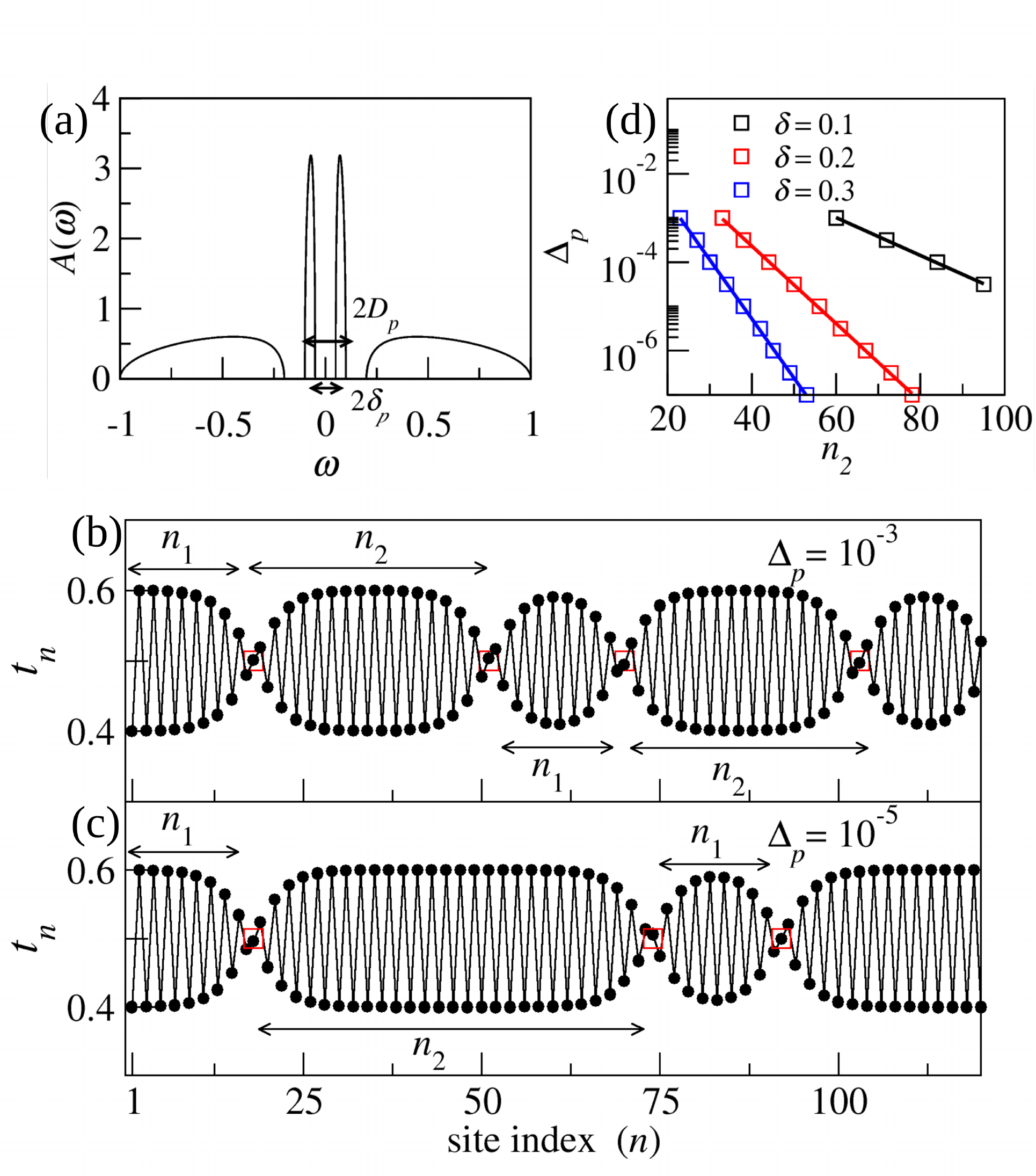}
  \caption{Beating in auxiliary chain hopping parameters, $\{t_n\}$ and its relation to the width of mid-gap sharp peaks: In panel (a), we design a toy model spectrum of full bandwidth $2D$ consisting of two outer SSH bands separated by a gap $2\delta$. The spectrum also consists of two inner SSH bands of full bandwidth $2D_p$ and gap $2\delta_p$. The width of these mid-gap peaks is defined as $\Delta_p$, such that $\Delta_p=D_p-\delta_p$. In panels (b) and (c) we plot the respective set of $\{t_n\}$ for $\Delta_p=10^{-3}$ (panel (b)) and $\Delta_p=10^{-5}$ (panel (c)), at a fixed $\delta=0.2$ and $D=1$. Clearly, the respective set of set of $\{t_n\}$ represents a periodic modulation of two SSH chains of domain lengths $n_1$ and $n_2$, with domain walls indicated as red squares. The width $(\Delta_p)$ of the inner SSH bands mimicking the effect of sharp mid-gap peaks is also a measure of the hybridisation energy between these domain walls as shown in panel (d), where we plot $\Delta_p$ as a function of $n_2$. The domain length, $n_2$ (squares) follows $\Delta_p\sim D \exp(-n_2 \delta/D)$ (fit represented as lines).}
  \label{fig:s2}
\end{figure}

In Fig~\ref{fig:s2}(a) we show such a toy spectrum of full bandwidth $2D$ consisting of two outer SSH bands separated by a gap $2\delta$. Inside this spectral gap of $2\delta$, there exists two inner SSH bands of full bandwidth $2D_p$ and gap $2\delta_p$, centred around $\omega_p$, such that the width of the inner SSH band is given by, $\Delta_p=D_p-\delta_p$. In order to understand the effect of the mid-gap features, we keep $\omega_p$ fixed and vary $\Delta_p$. We use the moment expansion technique, discussed in section~\ref{s3} to determine the respective tight binding chain represented by $\{t_n\}$ for $\Delta_p/D=10^{-3}$ (see Fig~\ref{fig:s2}(b)) and $\Delta_p/D=10^{-5}$ (see Fig~\ref{fig:s2}(c)) for a fixed $\delta/D=0.2,\,\omega_p/D=10^{-2}$. As seen in Figs~\ref{fig:s2}(b, c) the chain represented by $\{t_n\}$ consists of a periodic beat pattern with multiple domain walls (shown as red squares), such that each domain wall contributes to a topological localized state that hybridize amongst themselves. These periodic pattern and thereby the additional domain walls are completely absent in Fig~3(c, d) of the main text, where $\Delta_p=0$ because the spectrum consists of two mid-gap poles at $\pm\omega_p$. We may thus infer from Fig~3(c, d) of the main text that the location of the first domain wall and hence the length $n_1$ is pinned by the value of $\omega_p/D$. The additional beating is a manifestation of the presence of a band of topological states, represented as a mid-gap peak of width $\Delta_p$;  $\Delta_p$ determines the length $n_2$ that grows in size as $\Delta_p\to0$. In other words, the original SSH medium, denoted as $n_2$ is now interrupted by multiple domains of length $n_1$, due to the presence of a band of topological {\it defects} instead of just two topological excitations, such that the hybridisation $\Delta_p$ is determined by $n_2$ which is the real space separation between these {\it defects}. Indeed, these defects being topological in nature hybridise with each other via an exponentially localized $\Delta_p$ given by $\Delta_p\sim D \exp(-n_2\delta/D)$. This is shown in Fig.~\ref{s2}(d) where we vary $\Delta_p$ in our calculations and determine the respective $n_2$ (shown as squares) and observe that indeed $\Delta_p$ is exponentially localised in $n_2$ (solid lines).\\


\section{Boundary conductance of the auxiliary chain}
\label{s5}
A further signature of the topological nature of the Mott insulator self-energy, is a quantized fictitious $T=0$ `conductance' through the end of the auxiliary chain $G_{\rm{c}}/[2e^2/h]=\pi\Gamma A_{11}^{\rm{aux}}(\omega=0)=1$ (where $\Gamma$ is the hybridization to fictitious electrodes, and $A_{11}^{\rm{aux}}$ is the spectral function at the end of the electrode-coupled auxiliary chain). By contrast with the Mott insulator, the fictitious $T=0$ conductance through the end of the auxiliary chain precisely vanishes in the Fermi liquid phase, $G_{\rm{c}}/[2e^2/h]=\pi\Gamma A_{11}^{\rm{aux}}(\omega=0)=0$.


\section{PARTICLE-HOLE ASYMMETRY}
\label{sec:ph-asymm}
We now turn to the more complex situation away from particle-hole (\textit{ph}) symmetry. We quantify \textit{ph}-asymmetry according to the parameter $\eta=1-2\mu/U$, which involves the chemical potential $\mu$ and local Coulomb interaction strength $U$. For $\eta\ne 0 $,  the system is not \textit{ph}-symmetric, and we have $\text{Im}\Sigma(\omega)\ne \text{Im}\Sigma(-\omega)$. In the MI in particular, the Mott pole is not located precisely at the Fermi energy. These features lead to some differences in the mapping to the auxiliary chain and the subsequent analysis in terms of an emergent topology. However, the important conclusion, as explained below, is that the topological classification of the FL phase as trivial and the MI as topologically non-trivial carries over to the \textit{ph}-asymmetric case, and holds for the Mott transition arising at any $\eta$.

First, we should be careful to distinguish \textit{ph} asymmetry $\eta$, and the average filling $n$ which must be determined self-consistently. At \textit{ph} symmetry $\eta=0$, we have an exactly half-filled system, $n=1$. However, note that $n=1$ pertains throughout the Mott insulator phase for any $\eta$, and also $n\rightarrow 1$ as $U\rightarrow U_c^-$ in the metallic phase for any $\eta$.\cite{logan2015mott} The immediate vicinity of the Mott metal-insulator transition is in fact always at half-filling, for any $\eta$. 

In the metallic Fermi liquid (FL) phase, the lattice self-energy takes the usual FL form, $-t\text{Im}\Sigma(\omega)\sim (\omega/Z)^2$ at low energies $|\omega|\ll \omega_c$, where $\omega_c\sim Zt$ is the lattice coherence scale ($Z$ is the quasiparticle weight). This holds for \emph{any} asymmetry $\eta$, and as such there is an emergent low-energy \textit{ph} symmetry in the precise sense that $\text{Im}\Sigma(\omega)=\text{Im}\Sigma(-\omega)$ for all $|\omega|\ll \omega_c$ independent of $\eta$. Since low energies in $\Sigma(\omega)$ roughly correspond to large $n$ down the auxiliary chain, one therefore expects the `bulk' of the FL auxiliary chain to be the same as in the \textit{ph}-symmetric case already studied. At higher energies, \textit{ph} asymmetry shows up in an asymmetry between the upper and lower Hubbard bands; in general this leads to $\text{Re}\Sigma(\omega\rightarrow 0)\ne 0$ for $\eta\ne 0$. However, also note that $\text{Re}\Sigma(\omega\rightarrow 0) \rightarrow 0$ as $U\rightarrow U_c^-$ approaching the Mott transition from the FL side, independent of asymmetry $\eta$ and is as such a stronger definition of emergent \textit{ph} symmetry in the close vicinity of the Mott transition (see Ref.~\onlinecite{logan2015mott} for details). Aspects of the generic Mott transition at $\eta\ne 0$ might therefore be expected to be related to the \textit{ph}-symmetric case at  $\eta=0$.

We now discuss the generalization of the mapping to the auxiliary chain away from \textit{ph} symmetry. The auxiliary chain takes the form of Eq.~(2) of the main paper, with finite on-site potentials, $\{e_n\}$ in general. The CFE of the self-energy follows as,  
\begin{align}
    \Sigma(\omega)\equiv \Delta_0(\omega)=\cfrac{V^2}{\omega^+-e_1-\cfrac{t_1^2}{\omega^+-e_2-\cfrac{t_2^2}{\phantom{a}\ddots}}} \;.
    \label{eq:G_CFE_ph_asymm}
\end{align}
Similar to the discussion in Section~\ref{s1}, we can obtain $V^2=t_0^2$ from the relation $\int d\omega \mathcal{A}_0(\omega) \equiv -\tfrac{1}{\pi}\int d\omega~{\rm{Im}}\Delta_0(\omega) = t_0^2$; subsequently, for all $n>0$ one can recursively determine all $t_n$ and $e_n$ using the relations $\int d\omega \mathcal{A}_n(\omega) \equiv -\tfrac{1}{\pi}\int d\omega~{\rm{Im}}\Delta_n(\omega) = t_n^2$ and $-\tfrac{1}{\pi}\int d\omega~\frac{\omega{\rm{Im}}\Delta_{n-1}(\omega)}{t_{n-1}^2} = e_n$, where $e_1$ is the on-site energy of the boundary site in the isolated $H_{\text{aux}}$ coupled to the physical degrees of freedom via $H_{\text{hyb}}$. Note that in the {\it ph}-symmetric case, $e_n=0$ for all $n$.

\begin{figure}[htp!]
\includegraphics[clip=,scale=0.55]{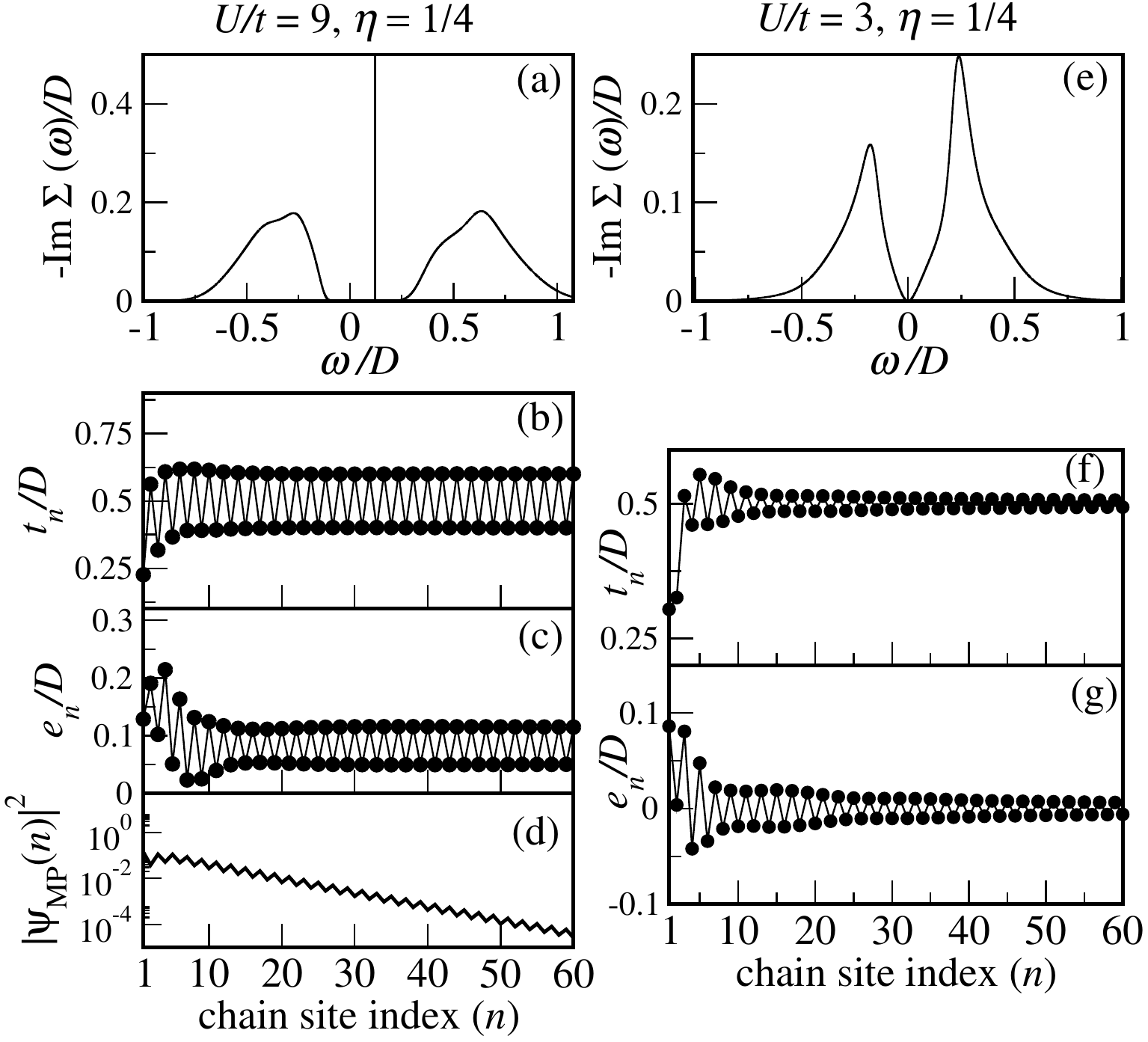}
  \caption{Lattice self-energy at $T=0$ and particle-hole asymmetry, $\eta=1/4$ obtained using NRG-DMFT [panels (a,e)] and corresponding auxiliary chain hopping parameters $\{t_n\}$ and onsite potential energies $\{e_n\}$ [panels (b,c) (MI), panels (f,g) (FL)]. Left panels show results for MI ($U/t=9.0,\,D=4.625$): the hard gap and pole at $E_{MP}\approx 0.12D$ produces a generalized SSH model in the topological phase, hosting a mode $\Psi_{\text{MP}}$ satisfying $H_{\text{aux}}\Psi_{\text{MP}}=E_{MP}\Psi_{\text{MP}}$ which is exponentially-localized on the boundary [panel (d)]. Right panels show results for FL ($U/t=3.0,\,D=4.5$). The low energy $\omega^2$ pseudogap in $\mathrm{Im}\Sigma$ results in an asymptotic $(-1)^n/n$ decay of the respective $t_n$ and $e_n$. Furthermore, the $\{e_n\}$ oscillate around zero (and tend to zero) at large $n$, indicative of a low energy emergent {\it ph}-symmetric dynamics.}
  \label{fig:MI_FL_ph_asymm}
\end{figure}

\subsection{Auxiliary chain representation of a {\it ph}-asymmetric Mott insulator}
\label{sec:ph_asymm_MI}
The self-energy $\text{Im}\Sigma(\omega)$ of a {\it ph}-asymmetric Mott insulator consists of two high energy Hubbard bands centred on some high energy positive $(\omega_+)$ and negative $(\omega_-)$ values respectively. There also exists an insulating charge gap between the Hubbard bands of width $2\delta=|\delta_+-\delta_-|$ where $\delta_\pm$ denotes the inner band edge on the positive (negative) energy side; here $|\delta_+|\ne|\delta_-|$ and $|\omega_+|\ne|\omega_-|$ unlike the {\it ph}-symmetric case. Additionally, in the {\it ph}-asymmetric case, the Mott pole inside the insulating gap is located away from the Fermi level at $\omega=\epsilon_0\equiv E_{\text{MP}}$. 
Thus the {\it ph}-asymmetric MI self energy is of the form $\Sigma(\omega)\equiv\Delta_0(\omega)=\Delta_0^{reg}(\omega)+\frac{\alpha_0}{\omega^+-\epsilon_0}$ where $\mathrm{Im}\Delta_0(\omega)=0$ for $\delta_-<\omega<\delta_+$. Since the Mott pole of weight $\alpha_0$ resides in the gap, we have $\delta_-<\epsilon_0<\delta_+$ and   
$\alpha_0^{-1}=-\frac{1}{\pi}\int_{-\infty}^{\infty}~d\omega\frac{\mathrm{Im}G(\omega)}{(\omega-\epsilon_0)^2}$, where $G(\omega)$ is the local lattice Green's function of the Mott insulator. 

For the CFE set up, we have, 
\begin{align}
    \Delta_{n-1}(\omega)&=t_{n-1}^2\tilde{G}_n(\omega),\\
    \tilde{G}_n(\omega)&=\frac{1}{\omega^+-e_n-\Delta_n(\omega)}.
\end{align}
Following the same logic based on the analytic structure of the complex $\Delta_n$'s as in the {\it ph}-symmetric case, we have 
\begin{align}
    \Delta_{2n-1}(\omega)&=\omega^+-e_{2n-1}-\frac{t_{2n-2}^2}{\Delta_{2n-2}^{reg}(\omega)+\frac{\alpha_{2n-2}}{\omega^+-\epsilon_{2n-2}}},\\
    \Delta_{2n}(\omega)&=\omega^+-e_{2n}-\frac{t_{2n-1}^2}{\Delta_{2n-1}(\omega)}\nonumber\\
    &=\Delta_{2n}^{reg}(\omega)+\frac{\alpha_{2n}}{\omega^+-\epsilon_{2n}},
\end{align}
where $n\ge1$ and every {\it even} -$\mathrm{Im}\Delta_{2n}(\omega)$ is gapped at low energies with a pole of weight $\alpha_{2n}$ located inside this gap at $\omega=\epsilon_{2n}$. Every {\it odd} -$\mathrm{Im}\Delta_{2n-1}(\omega)$ is gapped and regular and analytic in the complex plane. Since $\Delta_{2n-1}(\omega)$ is purely real for $\delta_-<\omega<\delta_+$ an isolated pole exists inside the gap in $\Delta_{2n}(\omega)$ at $\epsilon_{2n}$ where $\mathrm{Re}\Delta_{2n-1}|_{\omega=\epsilon_{2n}}=0$. 

The following relations are used to obtain $\{t_n\}$ and $\{e_n\}$ $\forall n\ge1$:
\begin{align}
    &e_n=-\int d\omega\frac{1}{\pi t_{n-1}^2}\omega\mathrm{Im}\Delta_{n-1}^{reg}(\omega)+\frac{\alpha_{n-1}\epsilon_{n-1}}{t_{n-1}^2},\\
    &t_n^2=-\int d\omega \frac{1}{\pi}\mathrm{Im}\Delta_{n}^{reg}(\omega)+\alpha_n,
\end{align}
where, $\mathrm{Im}\Delta_{n}^{reg}=\mathrm{Im}\Delta_{n}$ for {\it odd} $n$ and the respective pole weight $\alpha_n=0$ for all {\it odd} $n$. 
The pole weight, $\alpha_{2n}$ $\forall$ {\it even} sites is obtained using the relation,
\begin{align}
    \alpha_{2n}^{-1}=-\frac{1}{\pi t_{2n-1}^2}\int_{-\infty}^{\infty}~d\omega\frac{\mathrm{Im}\Delta_{2n-1}(\omega)}{(\omega-\epsilon_{2n})^2},
\end{align}
where, $\epsilon_{2n}$ is obtained by numerically locating the $\omega$ where $\mathrm{Re}\Delta_{2n-1}=0$ inside the gap. Since $\Delta_0(\omega)$ contains a pole inside the gap it is inevitable that $\Delta_2(\omega)$ will also contain a pole. However, unlike the {\it ph}-symmetric case the location of the pole will vary along the recursions, albeit remaining inside the gap, until the $\{t_n,e_n\}$ of the auxiliary chain settles down to a staggered alternating form without any attenuation [see Fig.~\ref{fig:MI_FL_ph_asymm}(b,c)].

It is important to note that delicate care must be taken with the numerical implementation of this scheme.



We illustrate the above mapping using the interacting self-energy obtained from NRG-DMFT for $U/t = 9$ and $\eta = 1/4$ shown in Fig.~\ref{fig:MI_FL_ph_asymm}(a). In Fig.~\ref{fig:MI_FL_ph_asymm} (b) we plot the auxiliary chain nearest-neighbour hopping parameters $\{t_n\}$, that displays a strictly alternating pattern akin to a standard SSH model and the {\it ph}-symmetric case in the bulk of the chain. Similar to the $\{t_n\}$ the on-site potentials, $\{e_n\}$ also oscillates alternating between two values, $\epsilon_A$ and $\epsilon_B$. This is plotted in Fig.~\ref{fig:MI_FL_ph_asymm} (c).

We now follow the analysis in Ref.~\onlinecite{li2014topological} and show that this auxiliary system is indeed topological. As shown in Fig.~\ref{fig:MI_FL_ph_asymm} (b,c) the bulk of the auxiliary chain in the Mott insulator has strictly alternating hoppings $t_A$ and $t_B$ as well as alternating
onsite potentials $\epsilon_A$ and $\epsilon_B$ on well-defined A and B sublattices. Fourier transforming yields $H =\int dk\vec{f}_k^{~\dagger} \mathcal{H}(k) \vec{f}_k$, where $\vec{f}_k^\dagger=(f_{Ak}^{~\dagger},f_{Bk}^\dagger)$ and, 
\begin{align}
	\mathcal{H}(k) = 
	\begin{pmatrix} 
	\epsilon_A & t_A + t_B e^{i k} \\ 
	t_A + t_B e^{-i k} & \epsilon_B 
	\end{pmatrix}.
\label{eq:nnn}
\end{align}
Following Ref.~\onlinecite{li2014topological} the coefficients of the Hamiltonian are re-parametrized as $t_A = t_0 (1 - \delta \cos\theta)$, $t_B = t_0 (1 + \delta \cos\theta)$, $\epsilon_{A} = h \cos(\theta + \phi)$, $\epsilon_{B}=h \cos(\theta - \phi)$. 

In the momentum space representation, the cyclic parameter $\theta$ plays the role of the momentum in a synthetic dimension alongside the usual quasimomentum $k$ to produce a two-dimensional effective Brillouin zone. The topological invariant is given by the Chern number $\text{Ch}= \frac{1}{2\pi}\int dk d\theta\left\lbrack \partial_\theta A_k - \partial_k A_\theta \right\rbrack$, where $A_{k}$ is the $k$ component of the Berry connection (defined in the usual way) and $A_\theta$ is the corresponding $\theta$ component. For $\theta \in [\frac{\pi}{2},\frac{3\pi}{2}]$ (as is the case for the MI self-energy shown here), the Chern number is explicitly given by $\text{Ch} = \frac12[\text{sgn}(2 h \sin\phi) - \text{sgn}(-2 h \sin\phi)]=\pm 1$ and the system is topological.
The respective parameters for our system, as obtained from the CFE mapping of the MI self energy for $U/t=9$ and $\eta=1/4$ plotted in Fig.~\ref{fig:MI_FL_ph_asymm} are $\epsilon_A = 0.24$ and $\epsilon_B = 0.52$, and hopping amplitudes $t_A = 1.85$ and $t_B = 2.78$. In order to cast this set of parameters in terms of $\delta$, $h$, $\theta$, and $\phi$ we utilise $t_0=2.31$ as obtained from the calculation and choose $\delta=0.5$ (which is a free parameter). This yields $\{h,\,\theta,\,\phi\}\approx\{-0.97,\, 1.98,\,-0.16\}$. The $k$ is chosen to be the momentum point where the band gap in $H_{\text{aux}}$ is minimum and is equal to the spectral gap in -Im$\Sigma$. This always occurs at $k=\pi$.
\begin{figure}[htp!]
    \includegraphics[clip=,scale=0.7]{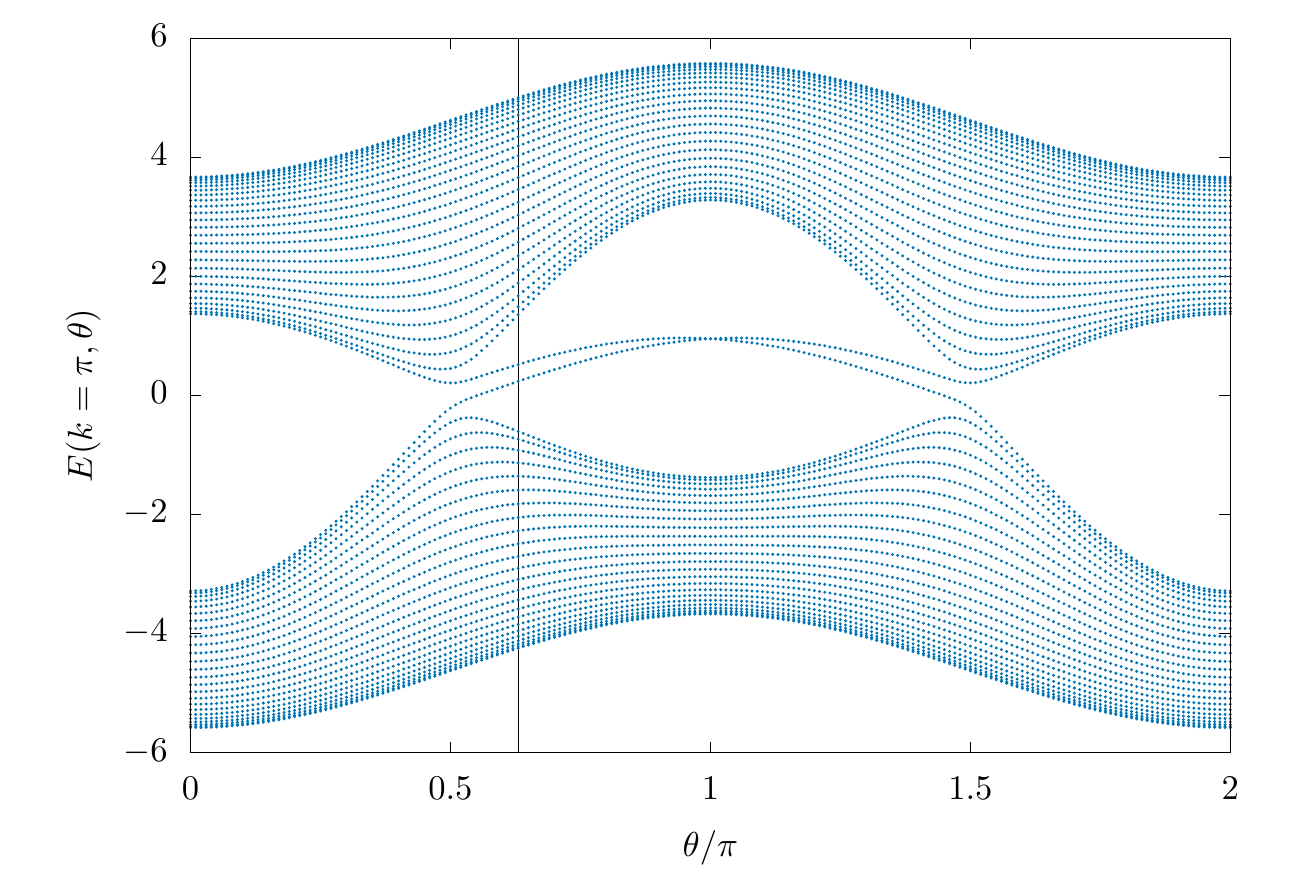}
    \caption{Band structure of the extended Li/Xu/Chen model introduced in Ref.~\onlinecite{li2014topological} in the ($E,\,\theta$)-plane at $k = \pi$, evaluated with parameters $\{h,\, t_0,\, \delta,\, \phi\}\approx\{-0.97,\, 2.31,\, 0.5,\, -0.16\}$ corresponding to the Mott insulator shown in Fig.~\ref{fig:MI_FL_ph_asymm}(left panel). The solid line marks the cut $\theta = 1.99$, which reproduces exactly the appropriate spectral function of the chain boundary (the two intragap bands correspond to localized states on the left and right boundaries; only the left boundary if physical in the semi-infinite auxiliary chain).}
    \label{fig:bands}
\end{figure}

Using the above parametrization, it follows that the Chern number $\text{Ch} = 1$ is quantized and the system is in the topological phase. The topological character of the parametrized Mott insulating phase shown here, is further exemplified by the intragap band crossing shown in Fig.~\ref{fig:bands}. Within this auxiliary model, the system is topological only if the intragap bands cross.\cite{li2014topological} This occurs if there exists a $\theta$ such that $h \sin\theta \sin\phi = 0$. This always occurs for $\theta \in [\frac\pi2,\frac{3\pi}{2}]$, which is when $t_A < t_B$, exactly as in the standard SSH model.  Therefore, the topological state is robust to perturbations in the hopping amplitudes as well as the on-site potentials.

Finally, we show that the mid-gap Mott pole (here arising at $E_{\text{MP}}\equiv e_0\approx0.12D$) corresponds to a bound state that is exponentially-localized on the boundary between the auxiliary system and the physical degrees of freedom of the lattice. We denote this `Mott pole state' 
$\Psi_{\text{MP}}$, and it satisfies $H_{\text{aux}}\Psi_{\text{MP}}=E_{MP}\Psi_{\text{MP}}$. 
Numerical diagonalization of $H_{\text{aux}}$ allows us to plot the wavefunction amplitude of this state as a function of chain site index $n$. This is plotted in Fig.~\ref{fig:MI_FL_ph_asymm}(d), and shows that $\Psi_{\text{MP}}$ is exponentially-localized on the edge of the chain. This is a result of the strict alternation of the chain parameters, with the coupling $t_1$ at the start of the chain being a \textit{weak} bond. Even though the chain parameters near the boundary exhibit variations/perturbations with respect to the bulk of the chain, the strict alternation guarantees that the Mott pole state is localized. 
This is confirmed by a generalization of the transfer matrix method, which yields explicitly $|\Psi_{\text{MP}}(n)|^2 \sim  \prod_{m=1}^{n-1} \frac{E_{\text{MP}} - e_{m} - t_{m-1}}{t_{m}}$. This expression gives a stringent condition connecting the parameters $t_n$, $e_n$ and the pole position $E_{\text{MP}}$ for a localized state. We see that the state has considerable robustness against perturbations to the chain parameters. In particular, if the Mott pole lies inside the gap (as it must in the insulator, by definition) then the corresponding state in the auxiliary chain is exponentially-localized on the boundary. Inverting the parity of the chain oscillations for all sites involves bulk gap closing; while local parity flips down the chain generate additional domain wall states (i.e.~multiple mid-gap poles, which are not seen in the Mott insulator). Although the Mott pole position, $E_{\text{MP}}$, is affected by the boundary potential, $e_1$, removing the boundary state by some boundary potential perturbation is equivalent to shifting the pole out of the gap -- in which case we no longer have a Mott insulator. When analyzing the robustness of the boundary-localized state to perturbations, we must therefore consider perturbations to the original Hubbard model, not unphysical perturbations to the auxiliary chain. Perturbations to the Hubbard model constitute  variations in the values $U/t$ and $\eta$.  However, provided such perturbations are not so drastic as to cause the system to cross into a different phase, the Mott insulator will always have a Mott pole inside the gap, and hence a corresponding boundary-localized state. We conclude that a Mott insulator, by definition, has an exponentially-localized state on the boundary of its auxiliary system, which is robust to the physically-relevant perturbations to the underlying Hubbard model.

We conclude that the auxiliary system is topologically non-trivial, even on breaking \textit{ph} symmetry.

\subsection{Auxiliary chain representation of a {\it ph}-asymmetric Fermi liquid}
\label{sec:ph_asymm_FL}
The FL form of the self-energy in the metallic phase of the \textit{ph}-asymmetric Hubbard model implies that $\Sigma(\omega)\equiv \Delta_0(\omega)\overset{\omega\to0}{\sim}a_0(\omega-\omega_0)+ib_0\omega^2$. The finite real part for $\omega\to0$ means that there is no pole in the odd $\Delta_n(\omega)$, unlike the {\it ph}-symmetric case studied above. This yields a simple recursion for the chain site hybridization functions, $\Delta_n(\omega)$, where $\Delta_n(\omega)$ $\forall n$ is a regular function. Thus, one can obtain the auxiliary chain representation for all $n\ge1$ via the recursion relation, $\Delta_n(\omega)=\omega^+-e_n-\frac{t_{n-1}^2}{\Delta_{n-1}(\omega)}$. All {\it even} $\mathrm{Im}\Delta_n$ will retain the low energy $\omega^2$ behavior. From our calculations based on the input FL self-energy for $U/t=3$ and $\eta=1/4$ we observed numerically that as the recursion proceeds, the {\it odd} $\Delta_n$ develop sharp features close to $\omega=0$. However, these features are not poles on top of a background function as for the {\it ph}-symmetric case. Thus, careful numerical evaluation with fine resolution in energy is required. 

In Fig.~\ref{fig:MI_FL_ph_asymm}(e) we plot -$\mathrm{Im}\Sigma(\omega)/D$ as a function of scaled frequency, $\omega/D$, where the effective bandwidth $D=4.5$ is the high energy cutoff chosen to obtain the auxiliary chain mapping for Hubbard model parameters, $U/t=3.0$ and $\eta=1/4$. Fig.~\ref{fig:MI_FL_ph_asymm}(f,g) shows the auxiliary chain parameters $\{t_n\}$ and $\{e_n\}$ in the metallic FL phase as per Fig.~2(e) of the main paper but now with $\eta=1/4$. The $\{t_n\}$ exhibit a similar alternating structure as for $\eta=0$, and in particular have the same $(-1)^n/n$ behaviour at large $n$, which is responsible for generating the low-energy FL pseudogap behaviour $-\mathrm{Im}\Sigma(\omega)\sim\omega^2$. The on-site potentials $e_n$ also show an alternation around zero energy, with the large-$n$ behaviour being damped by the same $1/n$ factor. The ``bulk'' of the chain in the FL phase therefore has $e_n\to 0$, consistent with the low-energy emergent {\it ph}-symmetry in the self-energy discussed above (and noting that low energies roughly correspond to large distances on the auxiliary chain). Note that there is no hard gap in the (metallic) FL phase. We have confirmed explicitly that all states of the auxiliary system in the FL phase are extended (none are boundary localized).\\

Hence we conclude that the FL phase is topologically trivial.\\


\section{Details of NRG-DMFT calculations}
\label{s6}
In this work, we solve the infinite-dimensional one-band Hubbard model on the Bethe lattice (Eq.~1 of the main text) numerically exactly at temperature $T=0$, using dynamical mean field theory\cite{georges1996dynamical} (DMFT), with Wilson's numerical renormalization group\cite{wilson1975renormalization,bulla2008numerical} (NRG) method as the underlying quantum impurity solver. NRG has the clear advantage over other impurity solvers (for this class of few-channel problems) in that real-frequency correlation functions can be obtained down to arbitrarily low energies, at any temperature including $T=0$, for any type of impurity interaction and strength.\cite{bulla1999zero} This is made possible by several important recent technical advances\cite{bulla1999zero,weichselbaum2007sum,pruschke2009energy,weichselbaum2012non,mitchell2014generalized,*stadler2016interleaved} over Wilson's original formulation of the method. Inhomogeneous systems can be treated, provided the local self-energy approximation remains valid.\cite{akm_localSE} Multi-band/orbital models and cluster extensions are now also within reach and could be analyzed within the framework presented here. 

For the purposes of the CFE of the self-energy, high-quality data for the self-energy are required. For the calculations presented in this paper, we use NRG discretization parameter $\Lambda=2$, retain $N_s=6000$ states at each iteration, and average the results of $N_z=20$ different bath discretizations. Total charge and spin projection quantum numbers are implemented. Correlation functions were obtained using the full density matrix method\cite{weichselbaum2007sum,anders2006spin} utilizing the complete Anders-Schiller basis. For converged solutions on the lattice, typically fewer than 20 DMFT iterations are required.


\vfill

%